\documentclass[]{aastex631}

\shorttitle{3-D haze distribution}
\shortauthors{Cohen et al.}
\begin{document}

\title{Haze optical depth in exoplanet atmospheres varies with rotation rate: Implications for observations}

\author[0000-0001-5014-4174]{Maureen Cohen}
\affiliation{School of GeoSciences,
The University of Edinburgh \\
Edinburgh, EH9 3FF, UK}
\affiliation{Centre for Exoplanet Science, The University of Edinburgh, UK}

\author[0000-0002-1487-0969]{Paul I. Palmer}
\affiliation{School of GeoSciences,
The University of Edinburgh \\
Edinburgh, EH9 3FF, UK}
\affiliation{Centre for Exoplanet Science, The University of Edinburgh, UK}

\author[0000-0001-6774-7430]{Adiv Paradise}
\affiliation{David A. Dunlap Department of Astronomy and Astrophysics,
University of Toronto \\
Ontario, Canada}

\author[0000-0001-7509-7650]{Massimo A. Bollasina}
\affiliation{School of GeoSciences,
The University of Edinburgh \\
Edinburgh, EH9 3FF, UK}

\author[0000-0001-7339-9495]{Paola Ines Tiranti}
\affiliation{School of Physics and Astronomy,
University of Leicester \\
University Road, Leicester LE1 7RH, UK}

\correspondingauthor{Maureen Cohen}
\email{s1144983@ed.ac.uk}



\begin{abstract}
Transmission spectroscopy supports the presence of uncharacterised, light-scattering and -absorbing hazes in the atmospheres of many exoplanets. The complexity of factors influencing the formation, 3-D transport, radiative impact, and removal of hazes makes it challenging to match theoretical models to the existing data. Our study simplifies these factors to focus on the interaction between planetary general circulation and haze distribution at the planetary limb. We use an intermediate complexity general circulation model, ExoPlaSim, to simulate idealised organic haze particles as radiatively active tracers in the atmospheres of tidally locked terrestrial planets for 32 rotation rates. We find three distinct 3-D spatial haze distributions, corresponding to three circulation regimes, each with a different haze profile at the limb. All regimes display significant terminator asymmetry. In our parameter space, super-Earth-sized planets with rotation periods greater than 13 days have the lowest haze optical depths at the terminator, supporting the choice of slower rotators as observing targets. 
\end{abstract}



\section{Introduction} \label{sec:intro}
Transmission spectroscopy promises insights into the atmospheres of a vast range of exoplanets, from the largest and hottest gas giants \cite{ehrenreich_nightside_2020} to sub-Earth-sized rocky worlds \cite{lustig-yaeger_detectability_2019}. Even before the launch and beginning of operations of the James Webb Space Telescope (JWST), attempts had been made to characterise exoplanet atmospheres with the Hubble Space Telescope \citep{damiano_transmission_2022, libby-roberts_featureless_2022, libby-roberts_featureless_2020, knutson_featureless_2014}, the Very Large Telescope \citep{bean_optical_2011}, and the Spitzer Space Telescope \citep{knutson_spitzer_2011}, with mixed results: the spectra were often featureless, allowing limited inferences to be made about the planet's atmospheric chemistry and composition \citep{kreidberg_clouds_2014}.  The greater sensitivity of the instruments on the JWST has partially overcome these limitations. For example, \cite{ahrer_identification_2023} reported direct detection of carbon dioxide in the atmosphere of WASP-39 b, previously only hinted at by Spitzer data \citep{wakeford_complete_2017}. On the other hand, \cite{lustig-yaeger_jwst_2023} reported a featureless transmission spectrum for LHS 475 b, a warm Earth-sized planet, potentially caused by a high cloud deck, a thin atmosphere, or no atmosphere. \cite{moran_high_2023} found a spectrum for warm super-Earth GJ 486 b that statistically diverged from flat but could not be fit by any atmospheric model. 

One cause of featureless transmission spectra that remains a challenge, including for the JWST and especially for observing Earth-like rocky planets, is the presence of clouds and hazes in a planet's atmosphere. An extensive literature exists on the topics of cloud and haze formation in exoplanet atmospheres and the impact they have on observations, including reviews by \cite{barstow_curse_2021} and \cite{helling_exoplanet_2019} for clouds and \cite{gao_aerosols_2021} for hazes. Theoretical predictions and laboratory work indicate that photochemically-produced hazes may be common in many types of exoplanets, including hot gas giants \cite{gao_aerosol_2020} and sub-Neptunes \citep{he_photochemical_2018, horst_haze_2018, kawashima_theoretical_2018, crossfield_trends_2017}, but also on Earth-like planets \citep{bergin_exoplanet_2023, arney_organic_2018, arney_pale_2017} and the early Earth itself \citep{arney_pale_2016, wolf_fractal_2010, haqq-misra_revised_2008, trainer_haze_2004}. In addition, observations of some planets have hinted at the existence of hazes due to the presence of a steep scattering slope at short wavelengths of the spectrum \cite{ohno_super-rayleigh_2020, pont_detection_2008} or, in the case of GJ 1214 b, a high Bond albedo of 0.51 combined with a featureless transmission spectrum \citep{kempton_reflective_2023}.

Theoretical studies of photochemical hazes on terrestrial planets have typically used the existing literature on organic hazes found on Titan, called tholins, as their starting point. For example, \cite{arney_pale_2017} used a 1-D coupled photochemical-climate model to simulate the effects of different stellar spectra on organic haze formation, using optical constants reported for Titan tholins by \cite{khare_optical_1984}. This study found that stellar spectra had a substantial impact on haze formation even within stellar classes: while the flaring M-dwarf (M3.5V) AD Leo produced only a thin haze layer at the top of the atmosphere, the quiescent M-dwarf (M4V) GJ 876 produced a haze layer 4-5 orders of magnitude thicker. As the simulations were 1-D and do not include transport by the planetary circulation, however, they may not represent the thickness of the haze layer at the planetary limb, the region of interest for transmission spectroscopy. \cite{steinrueck_3d_2021} and \cite{parmentier_3d_2013} studied the effect of the general circulation of hot Jupiters HD 189733 b and HD 209458 b on the gravitational settling of passive (i.e., not radiatively active) haze particles within the atmosphere. In their simulations, vertical mixing resulted in a relatively uniform abundance of small radius particles around the limb, but larger particles (with the size cut-off varying depending on the planet) displayed both spatial and temporal variability at the terminator. One-dimensional simulations are not able to capture this spatial variability.

The complexity of the formation and radiative impact of photochemical hazes\textemdash including coagulation and growth, sedimentation and transport, and effects on atmospheric heating profiles with consequent feedbacks on the circulation\textemdash make their representation in models a challenging task. In a recent work, \cite{steinrueck_photochemical_2023} simulated HD 189733 b with radiatively active tracers and haze optical properties from either soot-like particles or Titan-like tholins. They found that Titan-like organic hazes had a dramatic effect on the planetary general circulation, changing its structure sufficiently for the 3-D haze distribution to be very different from that simulated without the haze radiative effects. The soot-like hazes had no such impact. In parallel, a recent study by \cite{mak_3d_2023} used haze profiles from \cite{arney_pale_2017} to simulate radiatively active organic haze on the Archean Earth in a sophisticated 3-D general circulation model (GCM), the Met Office Unified Model. A notable finding is that organic haze could either cool or warm the planet depending on the thickness of the haze layer, further complicating the picture of haze-atmosphere coupled interactions. It is clear that while sophisticated 3-D haze studies such as these are urgently needed, faster models are also crucial to begin covering the vast parameter space represented by exoplanet hazes. 

In this paper, we present a fast, idealised 3-D GCM with radiatively active hazes to bridge the gap between 1-D models and complex 3-D GCMs. Our model, ExoPlaSim, is highly computationally efficient and was developed to map large parameter spaces and identify targets for simulation by more sophisticated GCMs \citep{haqq-misra_sparse_2022, paradise_exoplasim_2022}. Exoplanet GCMs inclusive of haze modelling were not available until very recently, but \cite{steinrueck_photochemical_2023}, \cite{mak_3d_2023}, and our ExoPlaSim extension now raise the prospect of applying a sampling protocol as described in \cite{haqq-misra_sparse_2022} to the study of exoplanet hazes. 

We use ExoPlaSim to perform a parameter space sweep of 3-D haze distribution on tidally locked terrestrial planets as a function of planetary rotation rate, with a focus on the terminator region. Although 1-D models of haze production and haze microphysics can examine how much haze may form at any location on a tidally locked planet, they do not account for horizontal transport from other haze-producing regions. The haze abundance at the planetary limb, where transmission studies probe, depends not only on local production, but also on horizontal transport from the dayside's substellar region, where photochemical haze production should peak. Our study thus aims to predict the terminator haze abundance for planets with different rotation periods given a fixed haze production amount on the dayside. The model and haze parameterisation as well as the numerical experiments are described in Section \ref{sec:methods}. In Section \ref{sec:results}, we present our results and in Section \ref{sec:discussion} we discuss their significance, potential conclusions for selecting observational targets, and the limitations of our study and planned further work. Conclusions are presented in Section \ref{sec:conclusion}.

\section{Methods} \label{sec:methods}

\subsection{Model} \label{subsec:exomodel}
We use the intermediate complexity GCM ExoPlaSim to perform our study. ExoPlaSim, documented in \cite{paradise_exoplasim_2022}, is based on the Planet Simulator \citep{lunkheit_planet_2011, fraedrich_planet_2005}, an Earth system model developed to maximise computational speed and portability. ExoPlaSim contains a spectral core and is optimised to run at T21 resolution (32 latitudes and 64 longitudes) with 10 vertical levels, which we maintain in our study. The model has a slab ocean and simulates moist atmospheric processes with vertical diffusion representing vertical mixing, Kuo-type deep convection \citep{kuo_further_1974, kuo_formation_1965}, Tiedtke shallow convection \citep{tiedtke_sensitivity_1983}, and dry convective adjustment. ExoPlaSim includes a radiative transfer module with two shortwave bands, a blue band from 0.3 to 0.75 $\mu$m and a red band from 0.75 to 14 $\mu$m, as well as one longwave band. The blue shortwave band ($\lambda < 0.75 \mu$m) represents Rayleigh scattering in the lowest atmospheric level, cloud scattering, water vapour absorption, and optional ozone absorption (not used in our study), while the red band ($\lambda > 0.75 \mu$m) includes water vapour absorption and both scattering and absorption by clouds \citep{lunkheit_planet_2011}. 

\cite{paradise_exoplasim_2022} modified the existing PlaSim source code to accommodate planets with a wide range of surface pressures, rotation rates, and stellar types, including tidally locked planets orbiting M-dwarf stars. At its standard T21 resolution, ExoPlaSim can compute a year of climate in roughly one minute of runtime \citep{paradise_gcm_2017}. ExoPlaSim has been used in previous studies of the habitability of Earth-like exoplanets \citep{checlair_no_2019, paradise_habitable_2019, checlair_no_2017, paradise_gcm_2017} and is particularly suited for simulations over long timescales \citep{chen_sporadic_2023} or large, unconstrained parameter spaces \citep{macdonald_climate_2022}. We refer to \cite{haqq-misra_sparse_2022} and \cite{paradise_exoplasim_2022} for detailed description of the model and its performance compared to other GCMs.

To represent haze particles circulating in the atmosphere, we used ExoPlaSim's built-in gridspace tracer transport module, supplemented by a gravitational settling scheme based on \cite{steinrueck_3d_2021} and \cite{parmentier_3d_2013}. The radiative effects of the haze are taken into account in a parameterisation for a scattering and absorbing aerosol given by \cite{stephens_radiation_1978}. We refer to Appendix \ref{Appendix} for a complete description of these parameterisations. With the haze module included, we were able to compute one year of model time in eight minutes when running on 32 processors. 

\subsection{Haze optical constants} \label{subsec:opticalconstants}
To simulate the effect of radiatively active hazes in the atmosphere, we use the complex refractive indices reported in \cite{he_optical_2023}. In this study, the authors generated haze analogs in the PHAZER chamber set-up of Johns Hopkins University \citep{he_carbon_2017} with an AC plasma energy source. The gas mixture used to approximate a water-rich 300 K atmosphere at 1000x solar metallicity was 56.0\% H$_2$O, 11.0\% CH$_4$, 10.0\% CO$_2$, 6.4\% N$_2$, 1.9\% H$_2$, and 14.7\% He. Although our simulated atmospheres of nitrogen gas with interactive water vapour and trace carbon dioxide and methane do not precisely replicate this gas mixture, the laboratory set-up and haze analogs produced in \cite{he_optical_2023} are a better fit for simulations of water-rich rocky planets than the previously widely used optical constants measured for organic tholins in a Titan-like atmosphere in \cite{khare_optical_1984}. We further choose to simulate this haze because the authors report both the complex refractive index and the density, a necessary input for the gravitational settling scheme, and because a large amount of haze was produced at 300 K, making it appropriate for simulations of hazy temperate planets. Ultimately, however, we expect photochemical haze compositions and production rates to be sensitive to the particular conditions on a planet and not easily generalised across atmospheres \citep{corrales_photochemical_2023, moran_chemistry_2020, he_laboratory_2018}. Our study examines the interaction between one potential kind of organic haze and the general circulation of tidally locked planets, without aiming to predict what kinds of organic haze may actually be present on temperate tidally locked M-dwarf planets.

Our simulations assume the haze particles are spherical and can be described using Mie theory \citep{mie_beitrage_1908}. We use \cite{he_optical_2023}'s complex refractive indices between 0.4 and 14 $\mu$m, binned to 10 nm, and the Python package MiePython \citep{prahl_miepython_2023} to calculate the wavelength-specific extinction, scattering, and backscattering efficiencies based on the size parameter derived from the particle radius. We use a particle of radius 500 nm based on findings from \cite{arney_pale_2017}, in which particles grew to this size for the quiescent M-class star GJ 876, and on the results of a sensitivity study of haze particle size reported in Section \ref{sec:results}. A pre-prepared BT-Settl-CIFIST stellar spectrum \citep{allard_bt-settl_2013}, also binned to 10 nm, is then used to compute the flux-weighted mean efficiencies for each shortwave band. Table \ref{tab:mieqs} lists the efficiencies for the two stellar spectra in our study for both shortwave bands, where band 1 indicates wavelengths $< 0.75 \mu$m and band 2 indicates wavelengths $> 0.75 \mu$m, as well as the stellar parameters. Comparison of the scattering and extinction efficiencies indicates that the particles are primarily scattering rather than absorbing.

\begin{table}[ht]
\centering
\begin{tabular}{|l|l|l|l|l|}
\hline
Star & TRAPPIST-1  & Wolf 1061  \\ \hline \hline
Q$_{ext}$ (band 1) & 3.42  & 3.00 \\ \hline
Q$_{scat}$ (band 1) & 3.24 & 2.79 \\ \hline
$\frac{dQ_{back}}{d\Omega}$ (band 1) (ster$^{-1}$) & 0.23 & 0.31 \\ \hline
Q$_{ext}$ (band 2) & 2.03 & 2.30 \\ \hline
Q$_{scat}$ (band 2) & 1.98 & 2.24 \\ \hline
$\frac{dQ_{back}}{d\Omega}$ (band 2) (ster$^{-1}$) & 0.03  & 0.04 \\ \hline
Stellar temperature (K) & 2709 & 3408 \\ \hline
Stellar metallicity (log$_{10}$ M/H) & 0.045 & -0.09 \\ \hline
log g (log$_{10}$ cm/s$^2$) & 5.2 & 4.9 \\ \hline
\end{tabular}
\caption{Mie extinction and scattering efficiencies and backscattering efficiency per steradian for two M-class stars for shortwave bands 1 ($0.3 < \lambda < 0.75 \mu$m) and 2 ($0.75 < \lambda < 14 \mu$m), as well as parameters defining the stellar spectra. Efficiencies are shown rounded to the second decimal place. Appendix \ref{Appendix} provides further details of the formulas used to calculate these quantities; please see Appendix Sections \ref{subsubsec:hazescheme} and \ref{subsec:miecalcs} for an explanation of the treatment of backscattering in particular.}
\label{tab:mieqs}
\end{table}%

\subsection{Source and sink} \label{subsec:sourcesink}
We position the haze source in the stratosphere at the top model level only in an analogy to Earth's Junge layer \citep{junge_world-wide_1961}. We assume the haze particles are photochemically produced \emph{in situ} by shortwave radiation and settle downwards to be redistributed by the tropospheric circulation. The source term is defined as:
\begin{equation}
    P = S_{top} \frac{SW_{top}}{SW_{max}}
\end{equation}
where $SW_{top}$ is the downward shortwave flux for each gridbox at the top level of the model and $SW_{max}$ is the maximum downward shortwave flux out of all gridboxes at the top level. $S_{top}$ is an arbitrary source coefficient in units of kg/kg. For a tidally locked planet, the source is at its maximum at the substellar point and falls off in all directions proportionally to the downward shortwave flux. The haze source can be dampened and reach an equilibrium of less than $S_{top}$ if, for example, the reflective effect of a haze layer reduces the incoming shortwave flux. We fix our source coefficient for all simulations at $10^{-7}$ kg/kg, based on haze profiles reported in \cite{arney_pale_2017} for an atmosphere with a methane to carbon dioxide ratio of 0.2 at an altitude of 25-30 km, the approximate height of our model top. A sink term is provided only in the bottom level. Its value is a constant $10^{-3}$ applied uniformly to the mass mixing ratio in the bottom model level (following \cite{steinrueck_3d_2021}), which is assumed to be a timescale for particles settling out of the atmosphere.

Table \ref{tab:freeparams} summarises the free parameters in the haze scheme, the values used in our simulations, and their sources.

\begin{table}[ht]
    \centering
    \begin{tabular}{|l|l|l|}
    \hline
        Parameter & Value & Source \\ \hline \hline
        Particle radius (nm) & 500 & \cite{arney_pale_2017} \\ \hline
        Particle density (kg/m$^3$) & 1262 & \cite{he_optical_2023} \\ \hline
        Source (kg/kg) & 10$^{-7}$ & \cite{arney_pale_2017} \\ \hline
        Sink ($\cdot$) & 10$^{-3}$ & \cite{steinrueck_3d_2021} \\ \hline
        Haze optical constants & Table \ref{tab:mieqs} & \cite{he_optical_2023} \\ \hline
    \end{tabular}
    \caption{Free parameters in the haze scheme, the values chosen for our simulations, and their sources.}
    \label{tab:freeparams}
\end{table}%

\subsection{Simulation set-up} \label{subsec:planets}
We performed a parameter space sweep of rotation rate for two idealised tidally locked aquaplanets, one representative of Earth-like terrestrial planets and one of super-Earths. For Earth-like planets, we use configuration values for TRAPPIST-1 e \citep{gillon_seven_2017}, an ideal candidate because it is used as a benchmark in the TRAPPIST-1 Habitable Atmosphere Intercomparison \citep{turbet_trappist-1_2022, sergeev_trappist-1_2022, fauchez_trappist-1_2022}. To extend our study to super-Earths, we use values for Wolf 1061 c \citep{wright_three_2016}. Table \ref{tab:planparams} summarises the parameters used in our simulations, which we take from the Arecibo Habitable Planets Catalog \citep{planetary_habitability_laboratory_phl_2023}. In addition to two size classes, our simulated planets represent two ends of the habitable temperature range, with ExoPlaSim simulating TRAPPIST-1 e as an eyeball planet (a sea ice-covered nightside and liquid water only in the substellar region) and Wolf 1061 c as a planet close to a runaway greenhouse state. Wolf 1061 c approaches 9\% atmospheric water vapour at the substellar point and is in a temperature and pressure regime that would lead to a runaway greenhouse in other GCMs (see \cite{haqq-misra_sparse_2022} for a comparison of ExoPlaSim and ExoCAM when simulating hot near-runaway atmospheres). However, our simulations remain numerically stable and reach an equilibrium with the values given in Table \ref{tab:planparams}, showing no further warming trend. As the failure to run away may be an artifact of the model's limitations with respect to water vapour treatment, these simulations should be seen as possible atmospheric states for warm temperate planets, rather than specific predictions for planets with the stellar constant of Wolf 1061 c. We discuss these limitations further in Section \ref{subsec:limitations}. The atmospheric composition of bulk nitrogen gas with interactive water vapour and trace methane and carbon dioxide in a 0.2 ratio is based on simulation parameters from \cite{arney_pale_2017} which allowed for the formation of a thick haze layer with a substantial effect on the climate in their study.

\begin{table}[ht]
\centering
\begin{tabular}{|l|l|l|l|l|}
\hline
Planet & TRAPPIST-1 e  & Wolf 1061 c \\ \hline \hline
Radius (R$_\bigoplus$) & 0.92 & 1.66 \\ \hline
Stellar constant (W/m$^2$) & 889 & 1777 \\ \hline
Gravity (m/s$^2$) & 9.1  & 12.1 \\ \hline
N$_2$ (bar) & 0.988 &  0.988 \\ \hline
CH$_4$ (bar) & 0.002 & 0.002 \\ \hline
CO$_2$ (bar) & 0.01  &  0.01 \\ \hline
\end{tabular}
\caption{Planet simulation parameters. Chemical abundances are given as partial pressures.}
\label{tab:planparams}
\end{table}%

Before carrying out the parameter space sweep, we ran each simulation for 75 model years to determine the time until radiative balance was reached at the top of the atmosphere. We found that while small net positive or negative energy anomalies continued to exist at top-of-atmosphere until ~60 years of model time, the temperature, atmospheric moisture, and haze profiles reached equilibrium and no longer showed an upward or downward trend after about 20 years. Accordingly, to save computational time, we run each simulation in the parameter space for 30 years and use the average of the final year of runtime for our analysis. All simulations are run with 1 bar ($10^5$ Pascal) of surface pressure at the standard T21 truncation (32 latitudes x 64 longitudes x 10 model levels) and initialised from a 250 K isothermal atmosphere. The top model level is located at ~10 mbar or ~25-28 km, representing the tropopause. We use a model timestep of 15 minutes.

\subsection{Study parameter space} 
\label{subsec:pspace}

To test the gravitational settling scheme, we first simulate a range of particle sizes and densities as passive tracers only, following \cite{steinrueck_3d_2021} and \cite{parmentier_3d_2013}. The particle radii include: 1 nm, 5 nm, 10 nm, 30 nm, 50 nm, 60 nm, 80 nm, 100 nm, 500 nm, and 1000 nm. These values cover a similar range as \cite{steinrueck_3d_2021}, but with greater focus on particles in the 10s of nanometers because laboratory experiments reported in \cite{he_photochemical_2018} found that haze particle size distributions tended to peak in the 30-80 nm range (although for a different gas composition than studied in \cite{he_optical_2023}). This set of simulations is repeated for particle densities of 1000 kg/m$^3$, 1262 kg/m$^3$, and 1328 kg/m$^3$. The latter two values are reported in \cite{he_optical_2023} as the densities of the organic haze analogues created in the study experiments at 300 K and 400 K, respectively. 1000 kg/m$^3$ is the reference value used in \cite{steinrueck_3d_2021} and \cite{parmentier_3d_2013} and approximates the density of black soot. The results of these size and density sensitivity tests are in line with those of \cite{steinrueck_3d_2021} and \cite{parmentier_3d_2013}. We briefly discuss them in Section \ref{sec:results}.

In our primary parameter space, we keep all inputs for each of the simulated planets fixed, but vary the rotation period from 1 to 30 Earth days, plus two additional simulations with periods of 6 and 12 hours. This gives a total of 32 simulations for each planet, or 64 in all. We use only one haze particle size and density: 500 nm, the size to which haze particles grew in the quiescent M-dwarf simulation in \cite{arney_pale_2017}, and 1262 kg/m$^3$, the density of haze analogues measured in \cite{he_optical_2023} for a water-rich 300 K atmosphere, together with the Mie efficiencies listed in Table \ref{tab:mieqs}. On Earth, particles of this size are in the accumulation phase, i.e. they accumulate in the atmosphere rather than settling out rapidly or agglomerating to form larger particles. In addition to these 64 simulations with radiatively active circulating haze, we perform another 64 simulations which are identical in all respects but do not include haze. We refer to this second set as ``control simulations'' and use them as a comparison to determine how the haze radiative feedback affects the surface temperature and general circulation in each simulation.

The aim of this experimental design is to study the potential distributions of organic hydrocarbon hazes on temperate tidally locked rocky planets as a function of rotation period (a known value for tidally locked planets). We focus particularly on the planetary limb, the region observed during transmission spectroscopy. While photochemical haze production may be expected to be highest in the substellar region, where stellar UV flux is the greatest, the haze abundance at the limb will depend on the general circulation. In this work, we do not vary the rotation period and stellar constant together, as we wish to isolate the effect of varying rotation. Instead, the inclusion of two planets with different stellar constants, stellar spectra, and sizes points at important factors in the larger parameter space affecting haze distribution which can be explored in future studies, while allowing one-to-one comparison between the planets based on rotation period. In the text hereafter, we refer to the set of simulations based on TRAPPIST-1 e planetary parameters as TRAP and those based on Wolf 1061 c as WOLF.

\section{Results}\label{sec:results}

Our simulations fall into three circulation regimes, previously described in the literature, and one transitional regime. Below we describe the general circulation and 3-D haze distribution of each regime, followed by an overview of the parameter space as a whole and a brief discussion of sensitivity to particle size and density. 

\subsection{Banded circulation regime}\label{subsec:reg1}
\begin{figure}
\begin{minipage}{0.99\textwidth}
\centering
\gridline{\fig{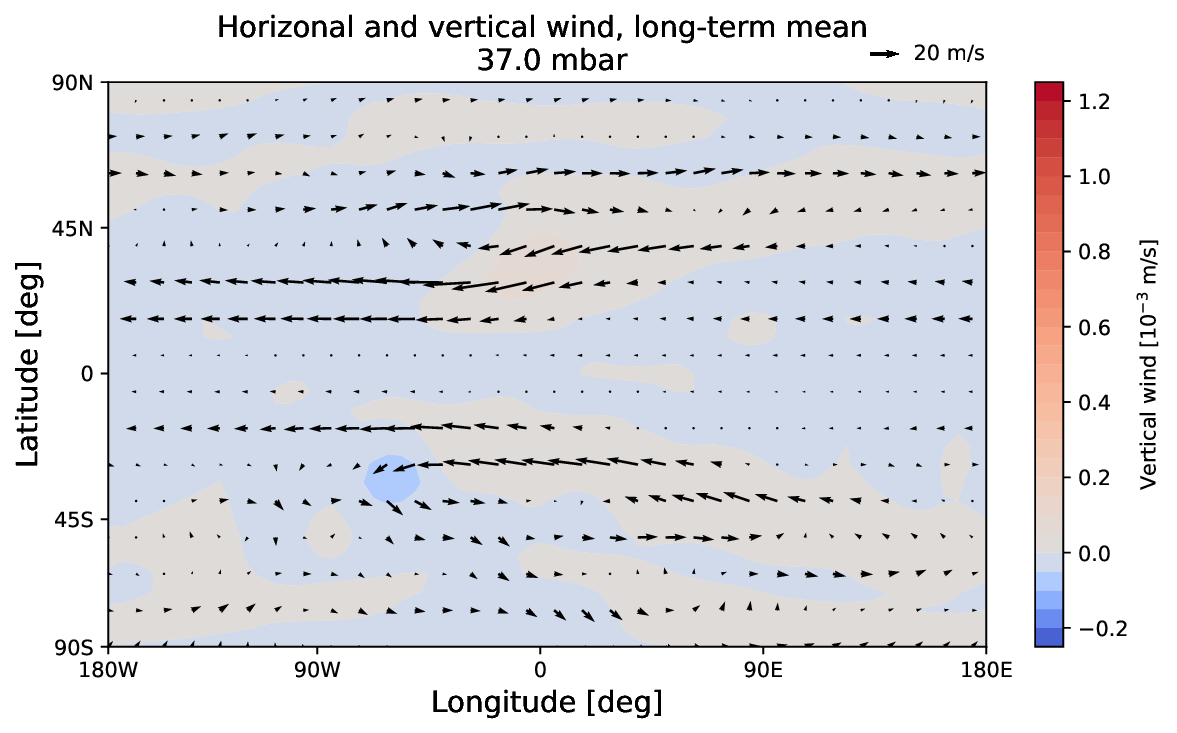}      {0.45\textwidth}{a) TRAPPIST-1 e, top of model }
          \fig{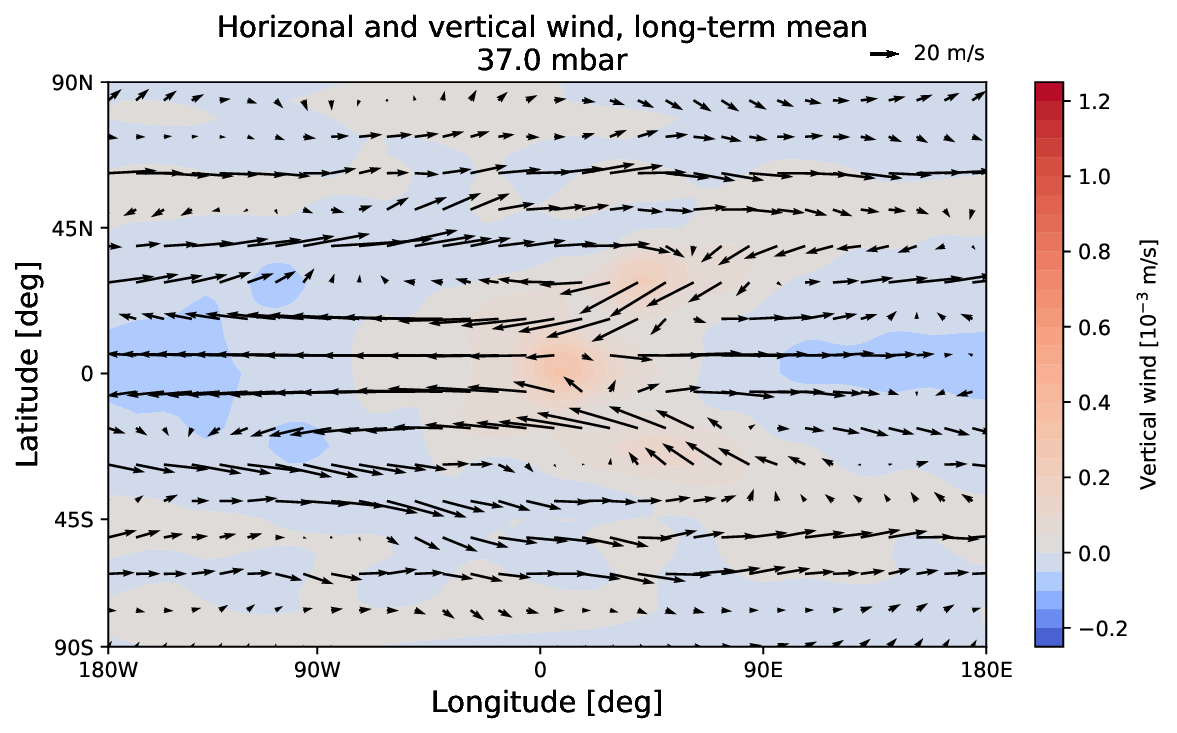}      {0.45\textwidth}{b) Wolf 1061 c, top of model }}
\gridline{\fig{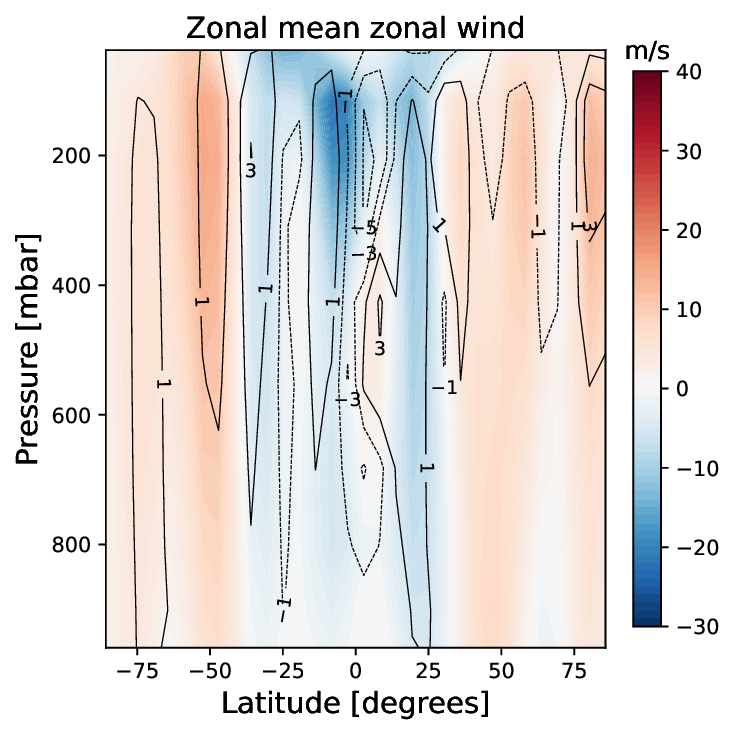}{0.33\textwidth}{c) TRAPPIST-1 e, zonal mean zonal wind}
          \fig{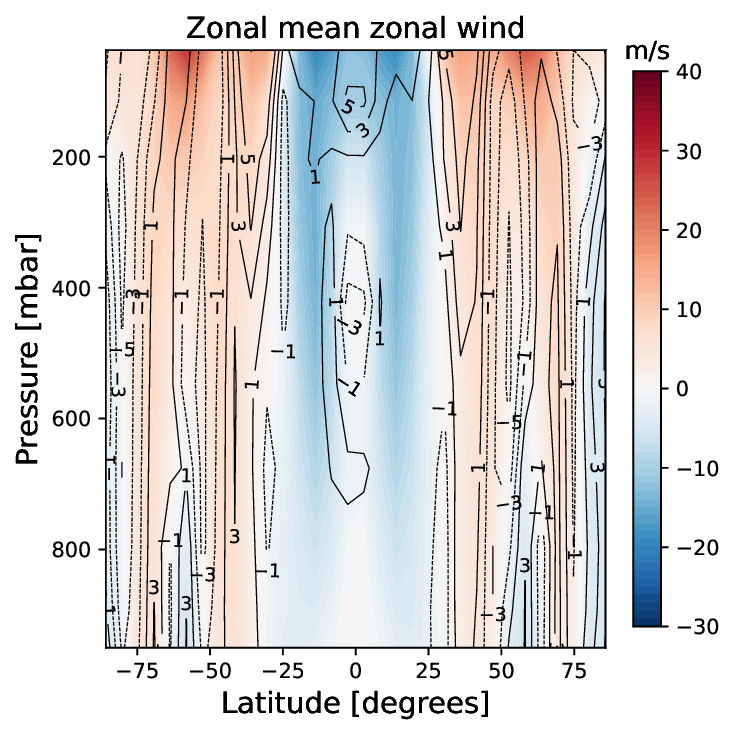}{0.33\textwidth}{d) Wolf 1061 c, zonal mean zonal wind}}
\caption{Top row: Horizontal and vertical winds in circulation regime 1 for a) TRAPPIST-1 e-type planet and b) Wolf 1061 c-type planet. Bottom row: Zonal mean zonal wind for c) TRAP and d) WOLF, with contour lines showing the difference in wind speed (absolute value) between the haze simulation and an identical simulation without haze (hazy minus control). Solid lines indicate positive values and dashed lines indicate negative. The rotation period shown is 0.25 days.}
\label{fig:reg1winds}
\end{minipage}%
\end{figure}

For rotation periods 0.25, 0.5, and 1 days, both TRAP and WOLF simulations fall into a banded circulation regime with eastward flow in the mid-latitudes and westward flow at the equator. This regime resembles the flow state for a planet with a 1-day rotation period shown in \cite{merlis_atmospheric_2010} (their Figure 6, top right), \cite{carone_connecting_2015} (their Figure 14 and 15), and \cite{noda_circulation_2017} (their Figure 11 b), as well as the ``rapid rotators'' described by \cite{haqq-misra_demarcating_2018}. We show the horizontal and vertical winds at the top model level in Figure \ref{fig:reg1winds} because this level is where the haze source is prescribed and where wind-haze interactions first take place. 

\begin{figure}
\begin{minipage}{0.99\textwidth}
\centering
\gridline{\fig{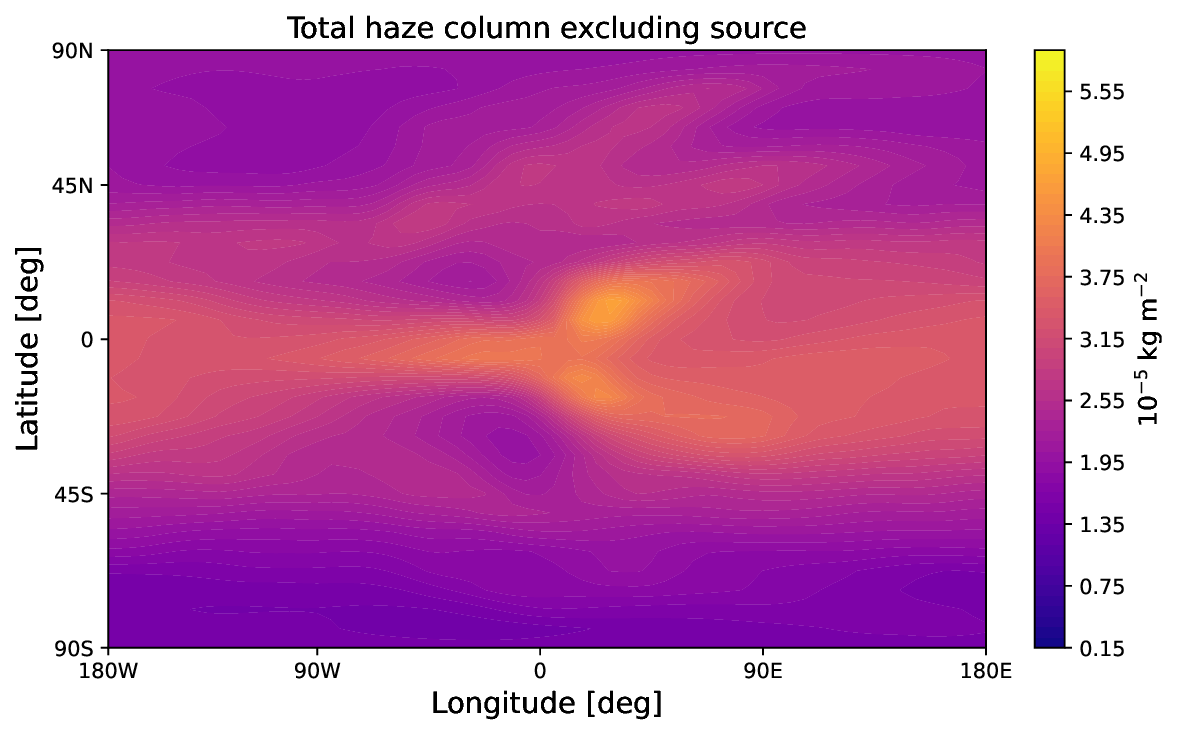}      {0.45\textwidth}{a) TRAPPIST-1 e}
          \fig{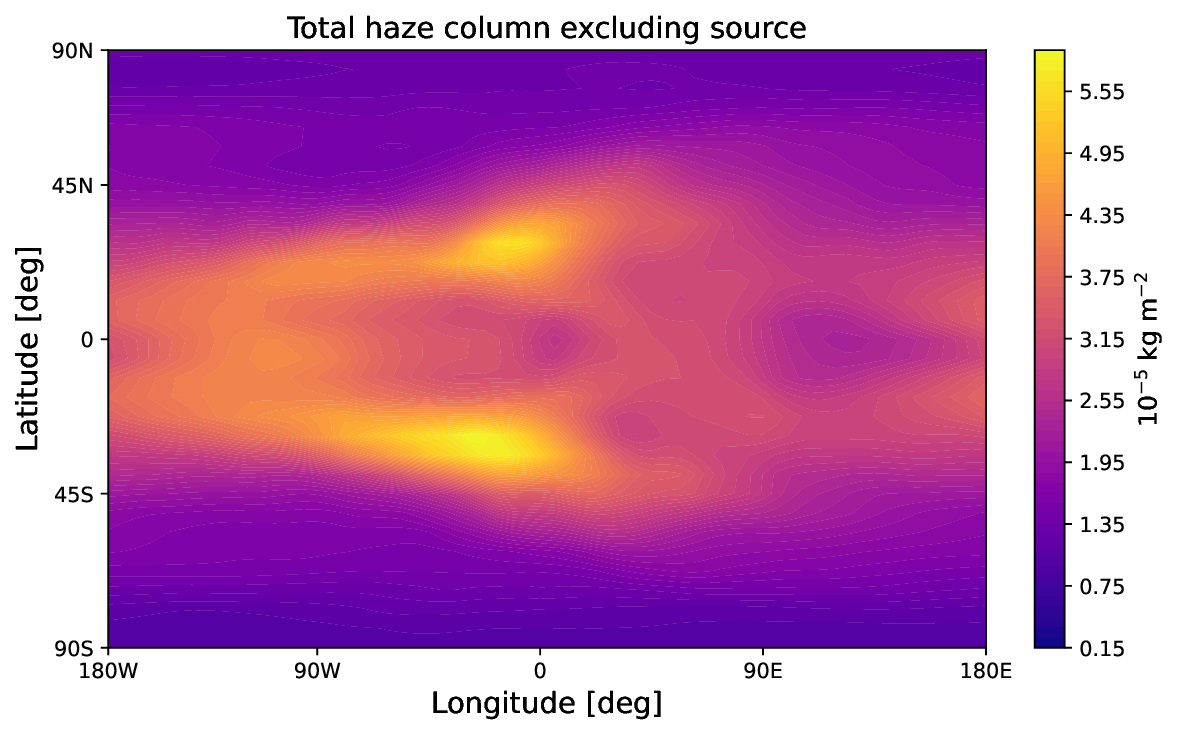}      {0.45\textwidth}{b) Wolf 1061 c}}
\gridline{\fig{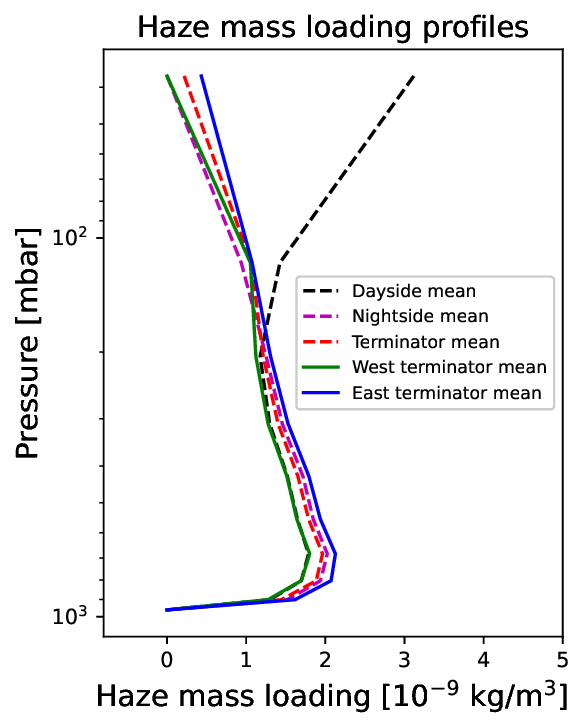}{0.33\textwidth}{c) TRAPPIST-1 e}
          \fig{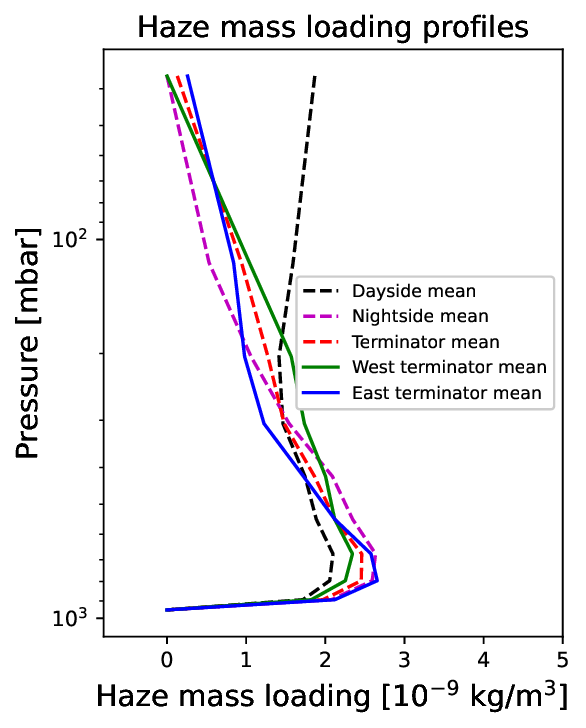}{0.33\textwidth}{d) Wolf 1061 c}}
\caption{Vertically integrated haze mass column for a) TRAP and b) WOLF and vertical haze mass profiles for c) TRAP and d) WOLF regime 1. The rotation period shown is 0.25 days.}
\label{fig:reg1mmr}
\end{minipage}%
%\end{figure}

%\begin{figure}[ht!]
\begin{minipage}{0.99\textwidth}
\centering
\gridline{\fig{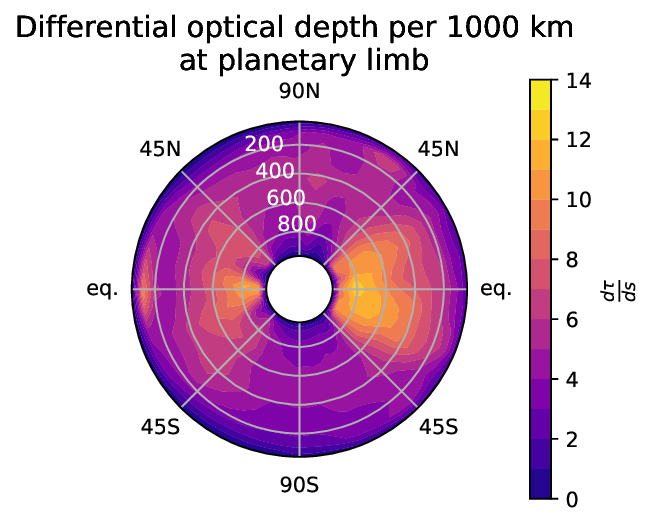}      {0.33\textwidth}{a) TRAPPIST-1 e}
          \fig{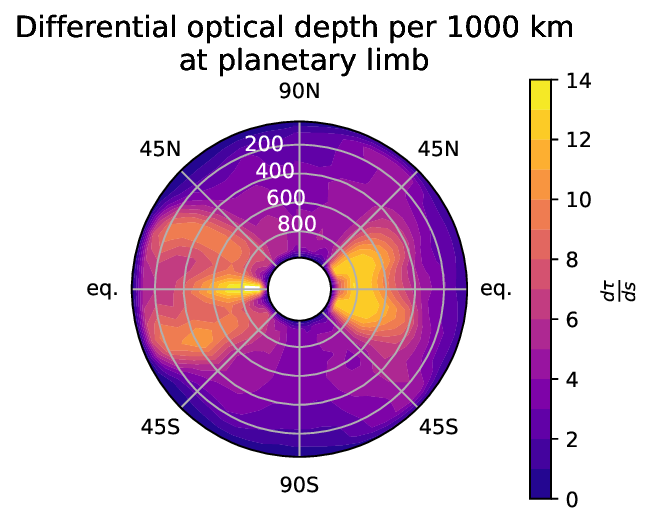}      {0.33\textwidth}{b) Wolf 1061 c}}
\caption{Differential optical depth per 1000 km in shortwave band 1 at the planetary limb for a) TRAP and b) WOLF for regime 1. The white text labels show pressure in mbar. The eastern (evening) terminator is shown on the right-hand side of each plot and the western (morning) terminator on the left-hand side of each plot to maintain consistency with other figures. The rotation period shown is 0.25 days.}
\label{fig:reg1tau}
\end{minipage}%
\end{figure}

Figure \ref{fig:reg1mmr} shows the vertically integrated haze mass column and the area- and density-weighted haze mass profiles for several regions on the planet: the dayside, nightside, terminator, and the western and eastern halves of the terminator split into separate areas. The TRAP case is comparatively well-mixed, with similar vertical haze profiles in all regions and a small gradient in the haze column from equator to pole. In contrast, the WOLF case is more spatially differentiated. Haze is concentrated closer to the equator and west of the substellar point, forming a distinct M- or W-shape. The vertical profiles reveal greater terminator asymmetry, with a hazier western terminator higher in the atmosphere, and a hazier eastern terminator below roughly 600 mbar.

In Figure \ref{fig:reg1tau}, we show an estimate of the haze differential optical depth in shortwave band 1 (blue) along the line of sight when the planet is viewed from the star. As a true 3-D treatment of the line of sight requires complex calculations and remains to be integrated into atmospheric retrievals \citep{wardenier_all_2022, caldas_effects_2019}, we use a fixed path length and assume the haze number density is constant along this path. Fig. \ref{fig:reg1tau} therefore represents a relative comparison of the extinction caused by the simulated TRAP and WOLF atmospheres rather than a quantitative prediction of optical depth. Like Fig. \ref{fig:reg1mmr}, Fig. \ref{fig:reg1tau} reveals more uniform haze mixing in the TRAP atmosphere both from equator to pole and from surface to top of the atmosphere, although there is still enhancement near the equator. In contrast, in the WOLF case haze remains largely concentrated within $\pm 40^\circ$. In addition, there is significant terminator asymmetry, with haze differential optical thickness decreasing above 400 mbar at the eastern terminator and remaining high to the model top at the western terminator.

\subsection{Transitional circulation regime}\label{subsec:reg2}
\begin{figure}
\centering
\gridline{\fig{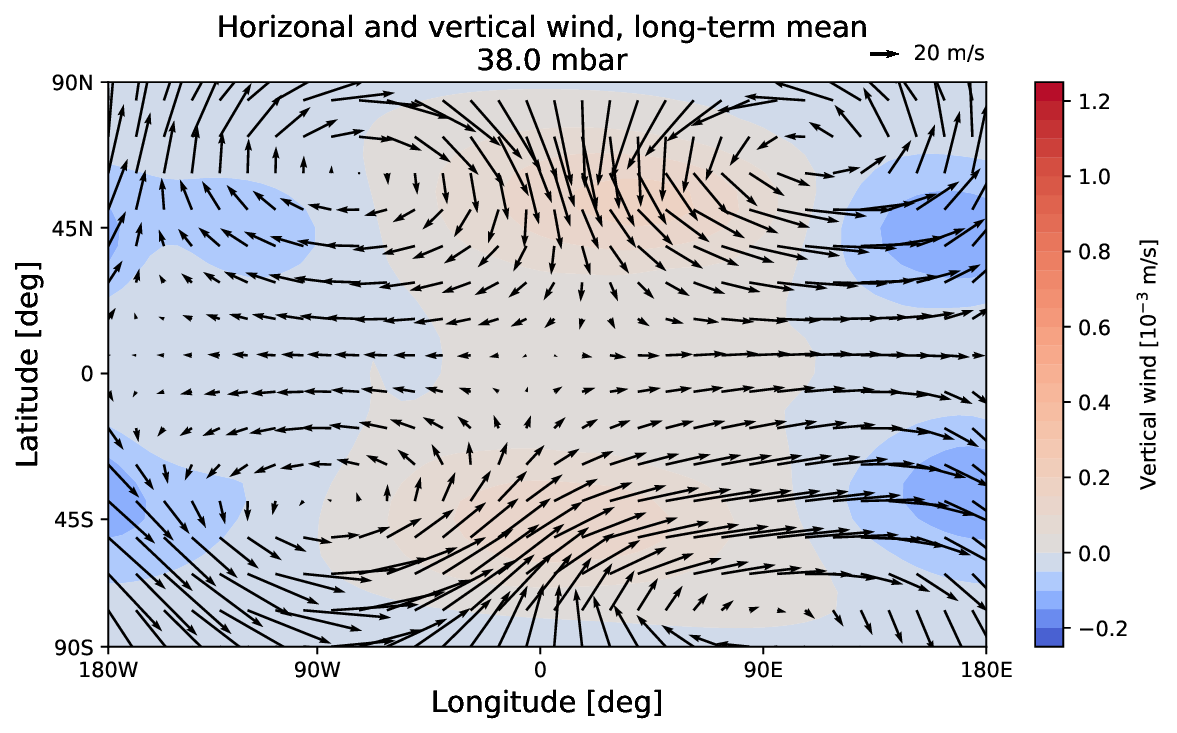}      {0.45\textwidth}{a) TRAPPIST-1 e, top of model }
          \fig{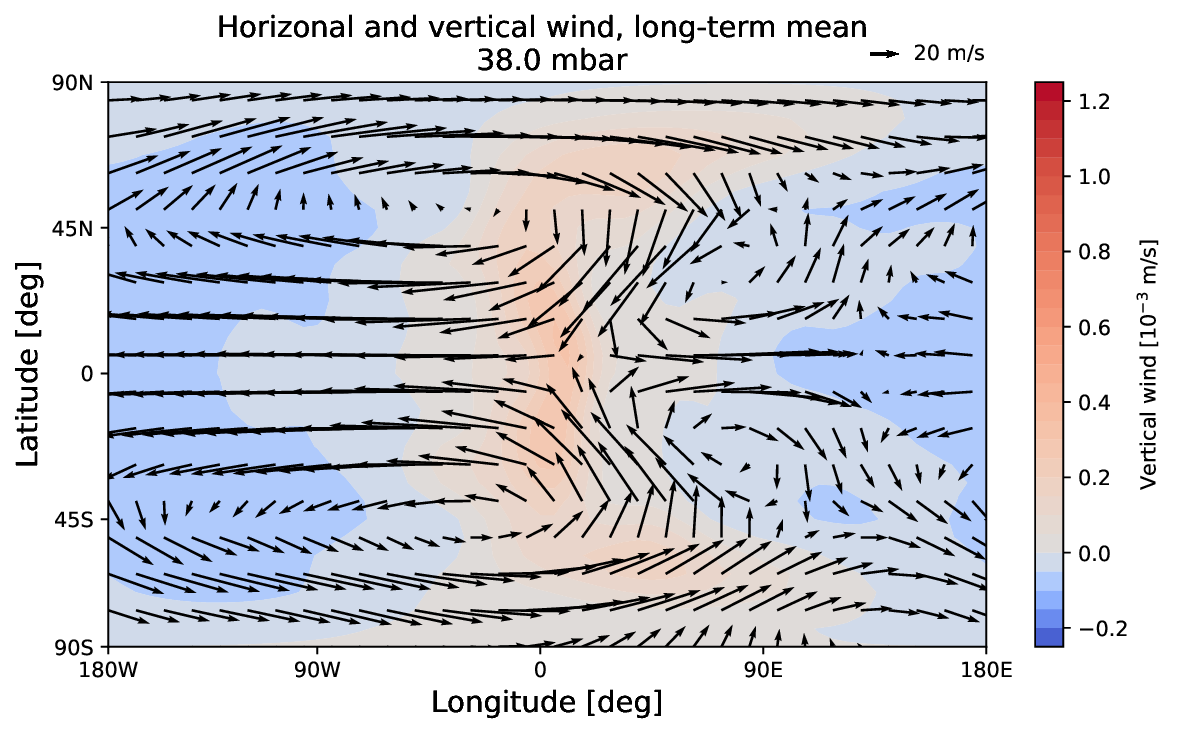}      {0.45\textwidth}{b) Wolf 1061 c, top of model }}
\gridline{\fig{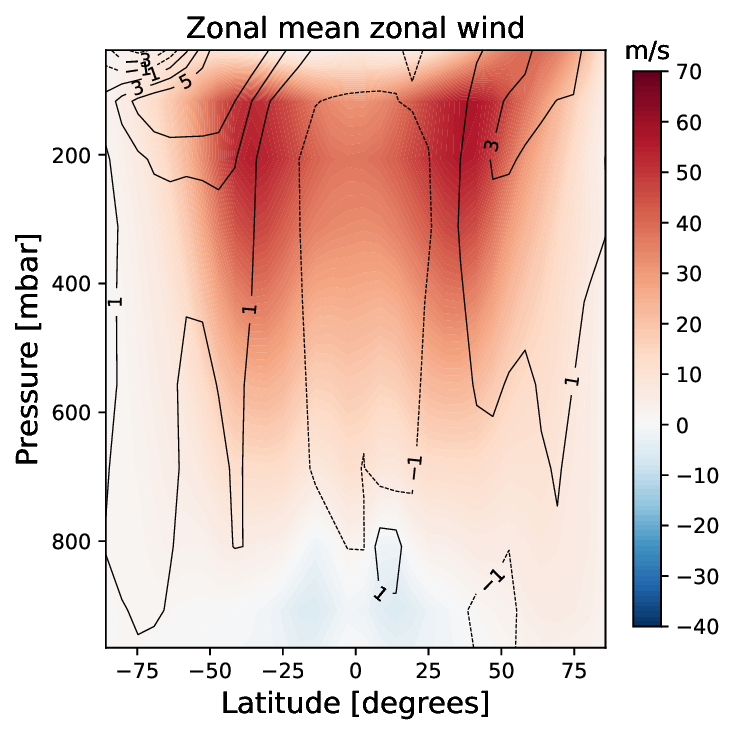}{0.33\textwidth}{c) TRAPPIST-1 e, zonal mean zonal wind}
          \fig{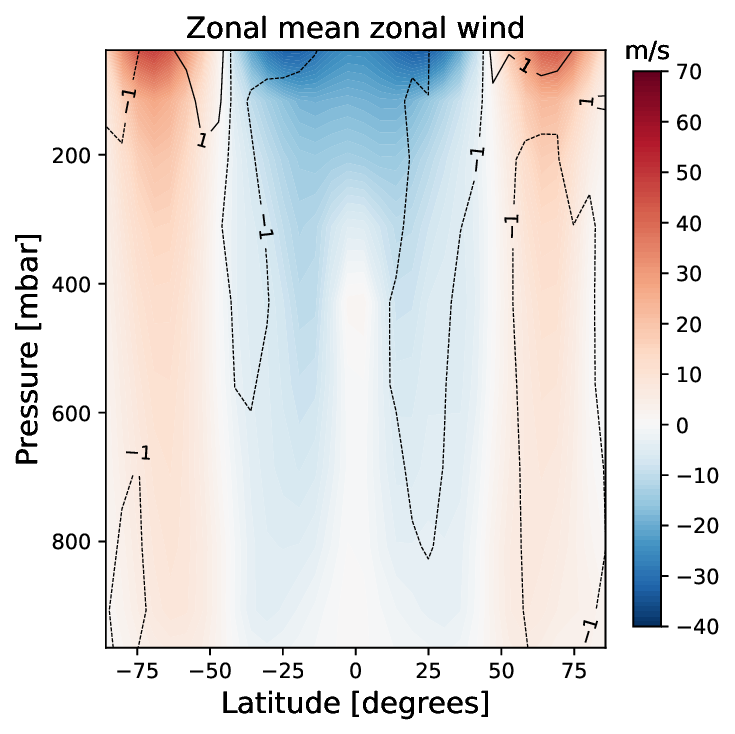}{0.33\textwidth}{d) Wolf 1061 c, zonal mean zonal wind}}
\caption{Top row: Horizontal and vertical winds in circulation regime 2 for a) TRAP and b) WOLF. Bottom row: Zonal mean zonal wind for c) TRAP and d) WOLF, with contour lines showing the difference in wind speed (absolute value) between the haze simulation and an identical simulation without haze (hazy minus control). Solid lines indicate positive values and dashed lines indicate negative. The rotation period shown is 2 days.}
\label{fig:reg2winds}
\end{figure}%

TRAP simulations with a rotation period of 2 days and WOLF simulations with periods of 2 and 3 days enter into a transitional space between circulation regimes characterised by a very low amount of haze. Figure \ref{fig:reg2winds} shows that the two models have different horizontal wind structures. In the TRAP case, there is a prominent quadrupolar pattern at the model top, while the WOLF case appears more similar to the banded regime, with higher wind speeds and the beginnings of gyre formation typical of the regime described in Section \ref{subsec:reg3} below. Similarly, the zonal mean zonal wind in TRAP no longer exhibits westward flow at the equator and instead forms an eastward jet with increasing wind speeds towards the mid-latitudes. In contrast, the zonal mean zonal wind in the WOLF case appears like a higher-speed, more regular version of the banded regime. However, both simulations show an increase in the magnitude and spatial extent of the upward wind on the dayside, with a corresponding increase in subsidence on the nightside.

\begin{figure}
\begin{minipage}{0.99\textwidth}
\centering
\gridline{\fig{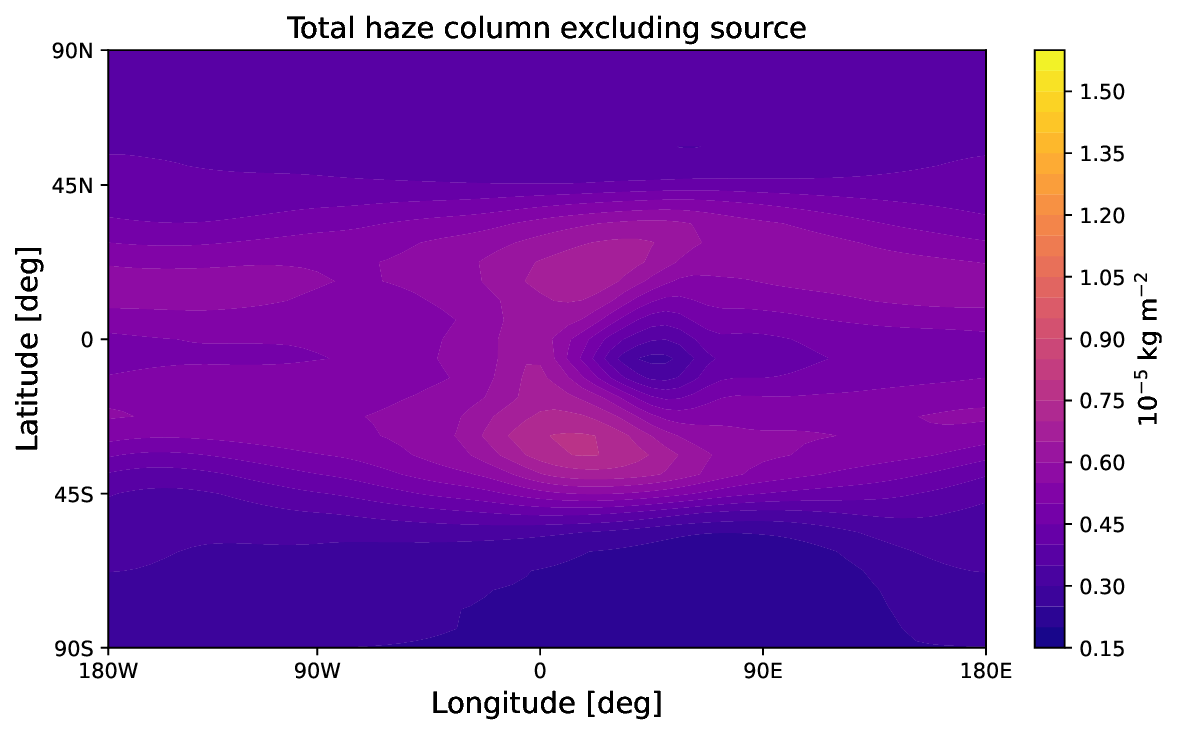}      {0.45\textwidth}{a) TRAPPIST-1 e}
          \fig{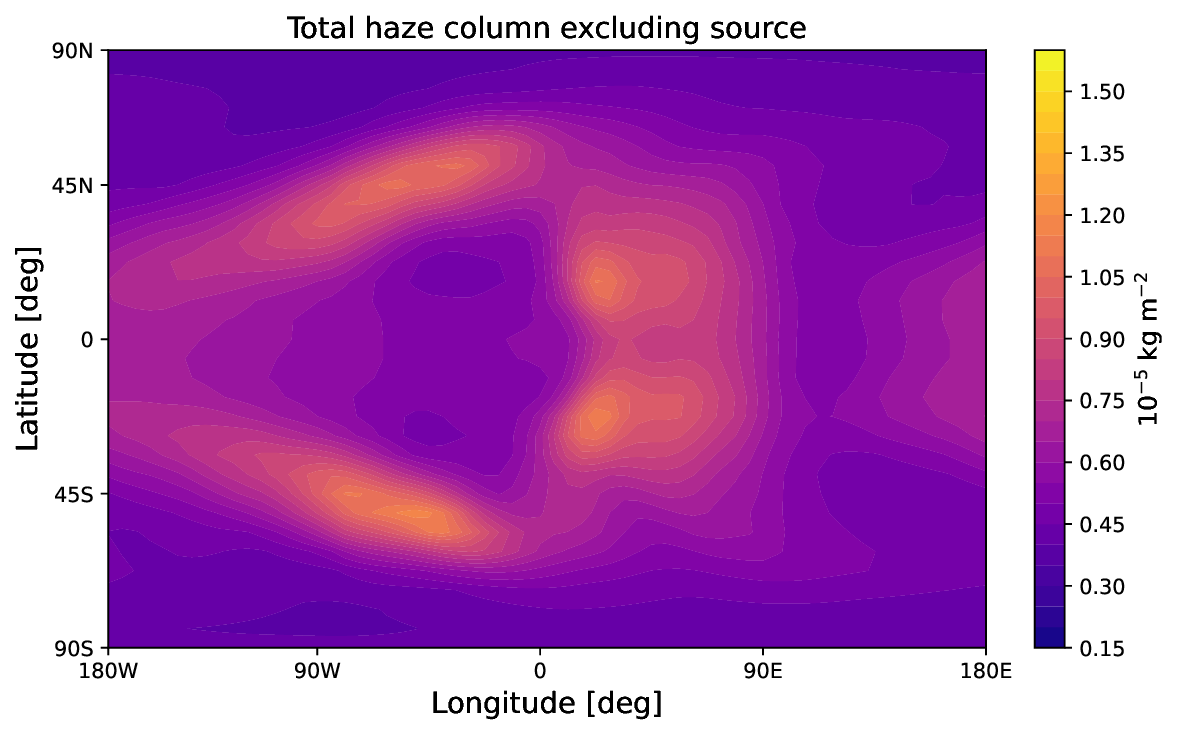}      {0.45\textwidth}{b) Wolf 1061 c}}
\gridline{\fig{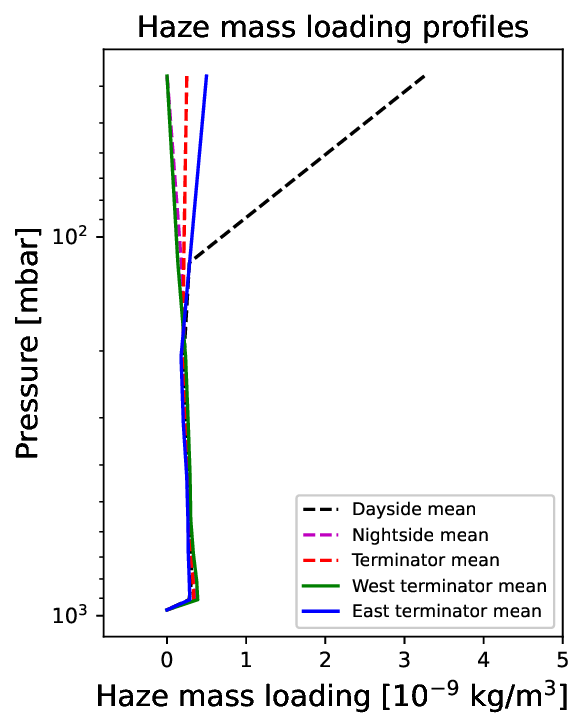}{0.33\textwidth}{c) TRAPPIST-1 e}
          \fig{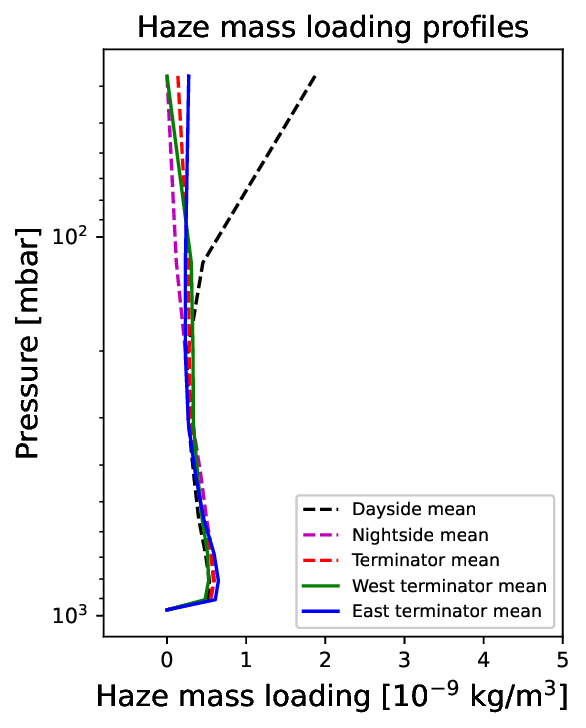}{0.33\textwidth}{d) Wolf 1061 c}}
\caption{Vertically integrated haze mass column for a) TRAP and b) WOLF and vertical haze mass profiles for c) TRAP and d) WOLF for regime 2. The rotation period shown is 2 days.}
\label{fig:reg2mmr}
\end{minipage}%
%\end{figure}

%\begin{figure}[ht!]
\begin{minipage}{0.99\textwidth}
\centering
\gridline{\fig{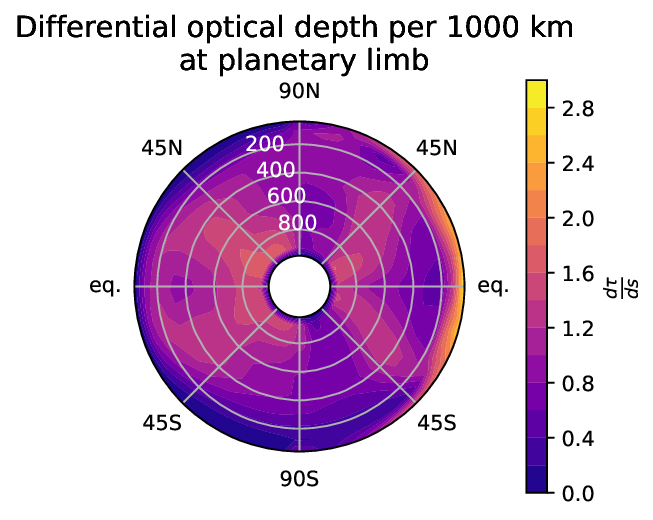}      {0.33\textwidth}{a) TRAPPIST-1 e}
          \fig{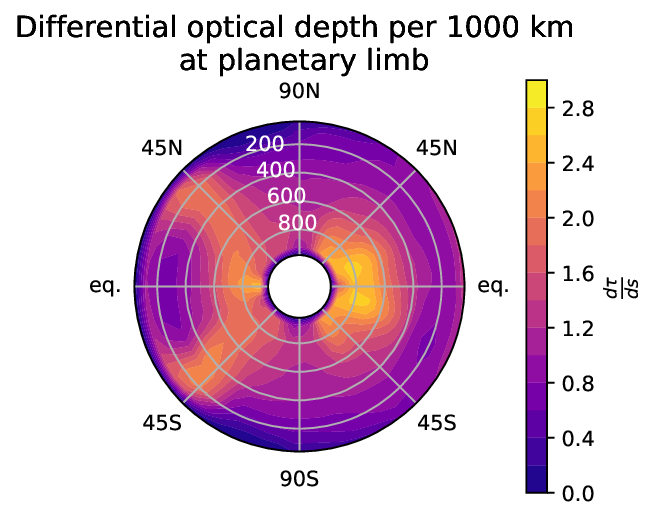}      {0.33\textwidth}{b) Wolf 1061 c}}
\caption{Differential optical depth per 1000 km in shortwave band 1 at the planetary limb for a) TRAP and b) WOLF regime 2. The eastern (evening) terminator is shown on the right-hand side of each plot and the western (morning) terminator on the left-hand side of each plot to maintain consistency with other figures. The white text labels show pressure in mbar. The rotation period shown is 2 days.}
\label{fig:reg2tau}
\end{minipage}%
\end{figure}

Figure \ref{fig:reg2mmr} reveals a low amount of haze throughout the atmosphere, with vertical profiles consistent across the planet at around $0.5 \times 10^{-9}$ kg/m$^3$ for both WOLF and TRAP. The vertical haze columns and differential optical depths shown in Figure \ref{fig:reg2tau} are about 25 percent of those in the banded regime. The low haze amount could be caused by, for example, more rapid settling of particles at the planetary surface, or a failure of particles to escape the source at the top level. The larger extent of the area of upward wind shown in Fig. \ref{fig:reg2winds} a) and b) and uniform lack of haze throughout the atmosphere in Fig. \ref{fig:reg2mmr} suggest the latter is more likely. A detailed analysis of the particle trajectories would be helpful to clarify this aspect, but is beyond the scope of this study. 

\subsection{Double jet circulation regime}\label{subsec:reg3}
\begin{figure}
\centering
\gridline{\fig{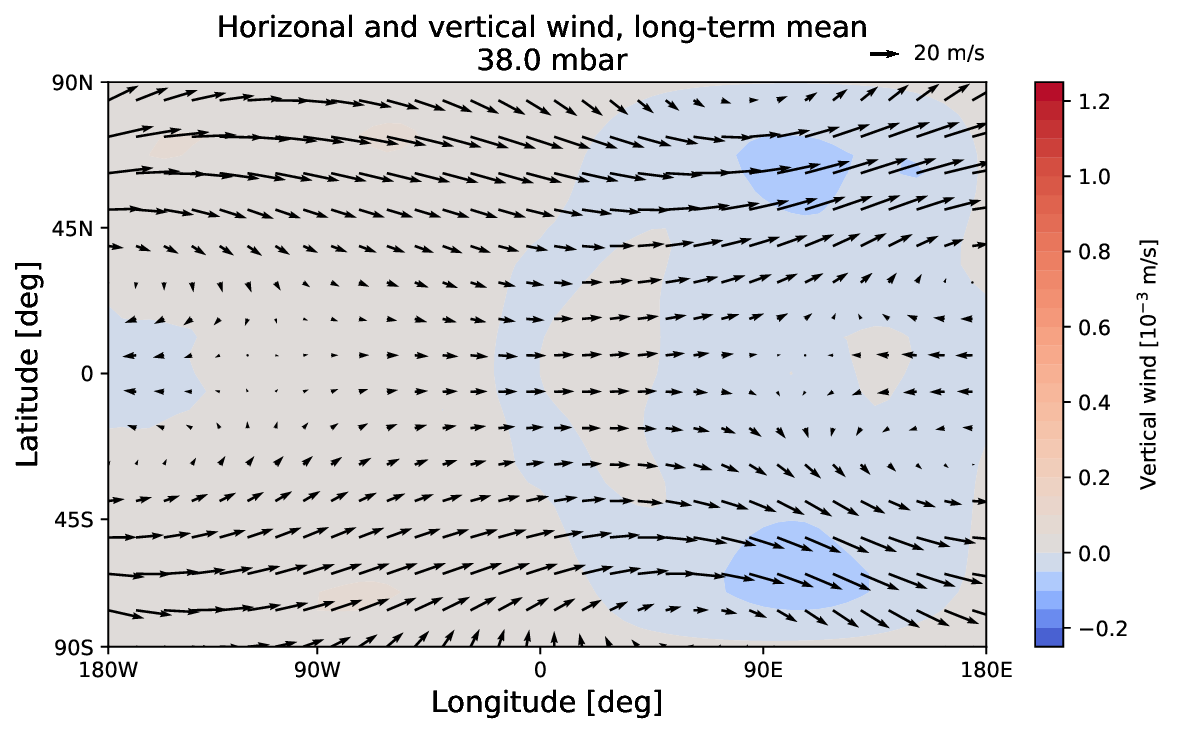}      {0.45\textwidth}{a) TRAPPIST-1 e, top of model }
          \fig{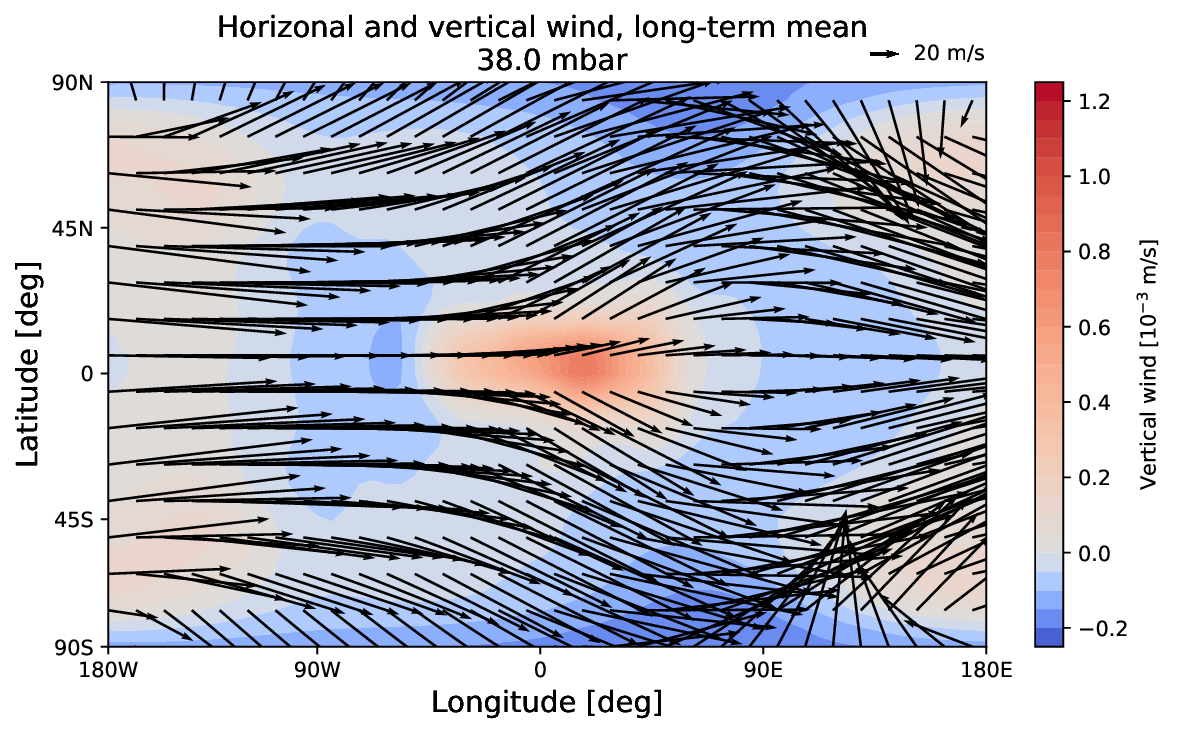}      {0.45\textwidth}{b) Wolf 1061 c, top of model }}
\gridline{\fig{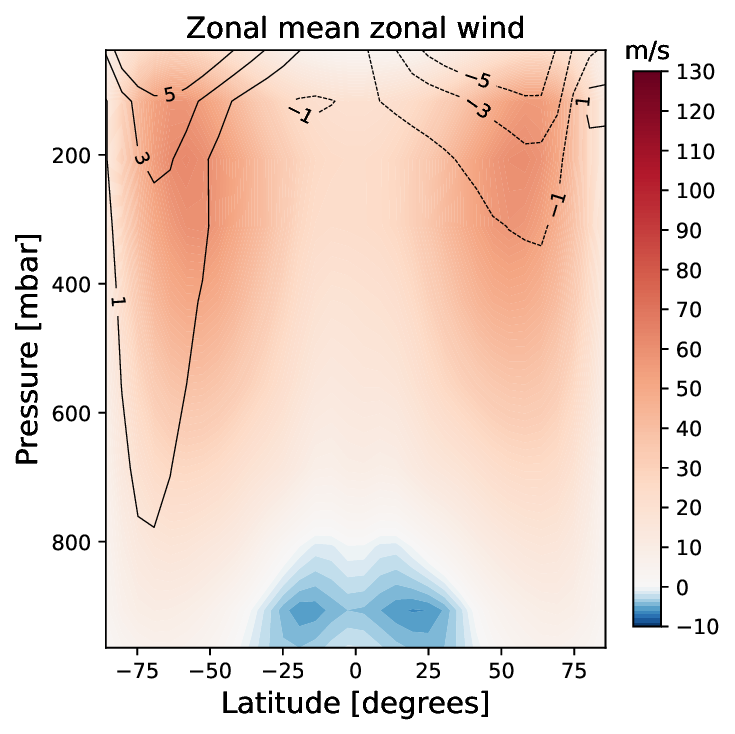}{0.33\textwidth}{c) TRAPPIST-1 e, zonal mean zonal wind}
          \fig{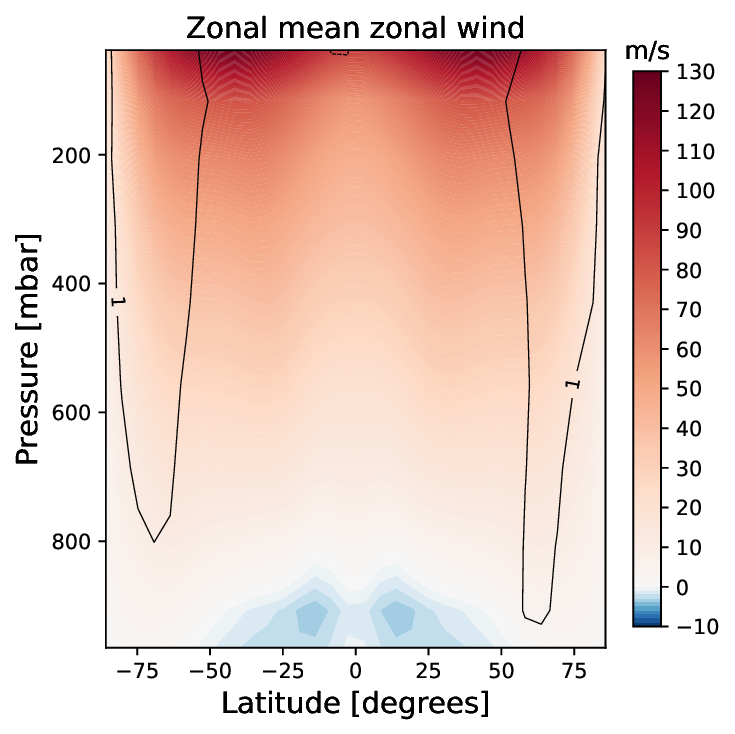}{0.33\textwidth}{d) Wolf 1061 c, zonal mean zonal wind}}
\caption{Top row: Horizontal and vertical winds in circulation regime 3 for a) TRAP and b) WOLF. Bottom row: Zonal mean zonal wind for c) TRAP and d) WOLF, with contour lines showing the difference in wind speed (absolute value) between the haze simulation and an identical simulation without haze (hazy minus control). Solid lines indicate positive values and dashed lines indicate negative. The rotation period shown is 6 days.}
\label{fig:reg3winds}
\end{figure}
\begin{figure}
\begin{minipage}{0.99\textwidth}
\epsscale{0.75}
\centering
\gridline{\fig{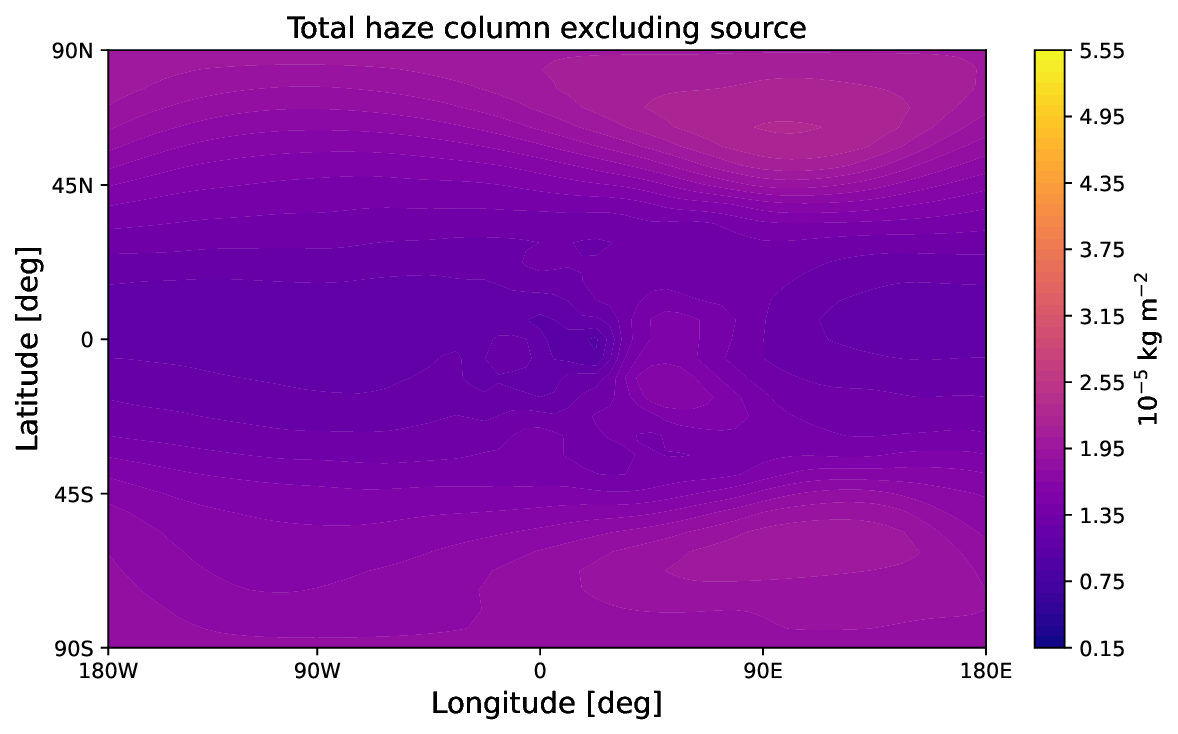}      {0.45\textwidth}{a) TRAPPIST-1 e}
          \fig{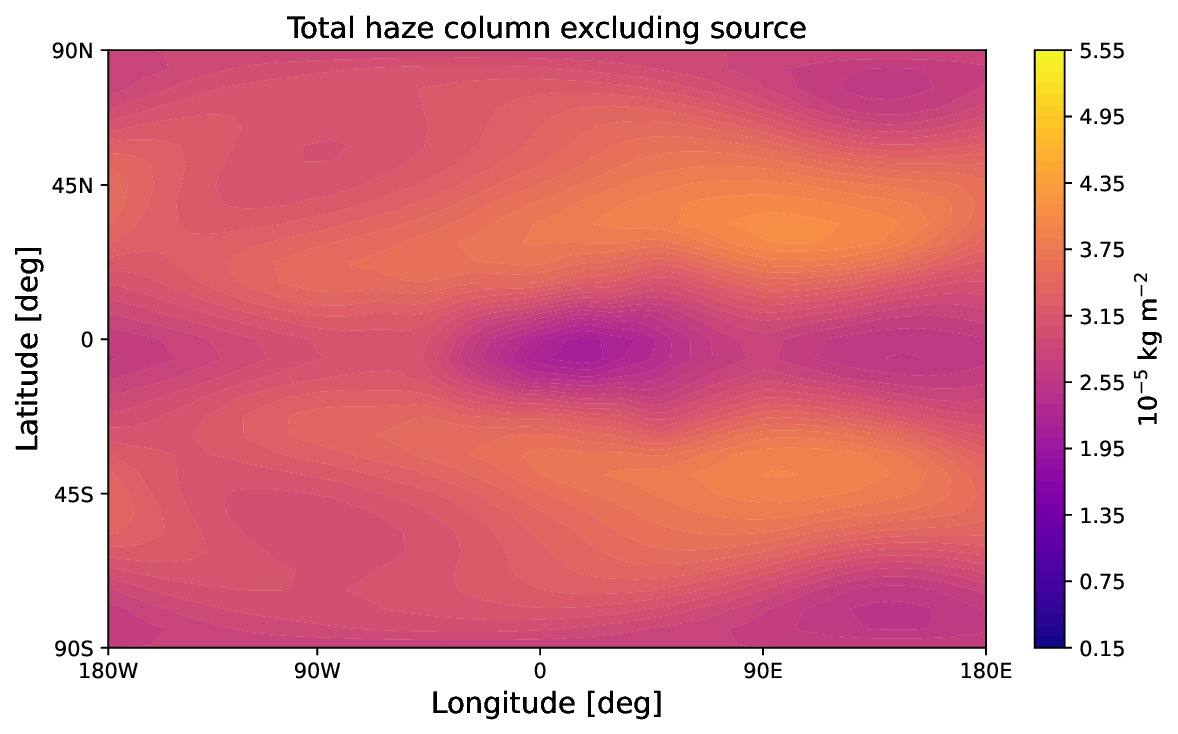}      {0.45\textwidth}{b) Wolf 1061 c}}
\gridline{\fig{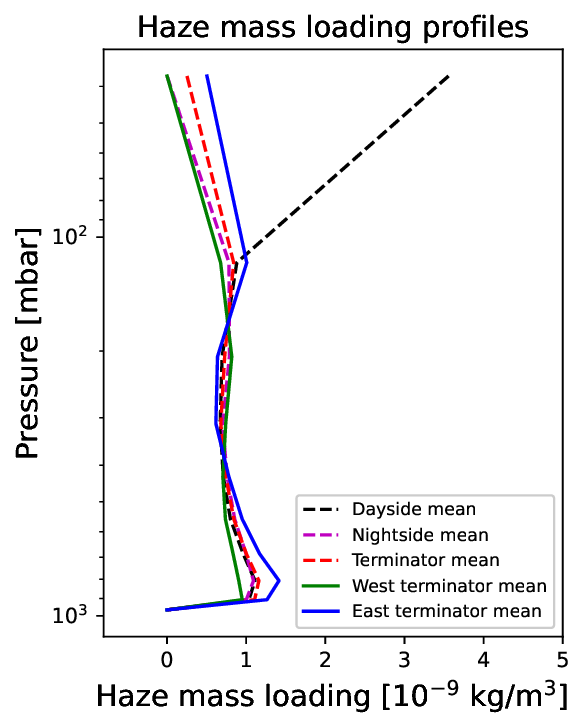}{0.33\textwidth}{c) TRAPPIST-1 e}
          \fig{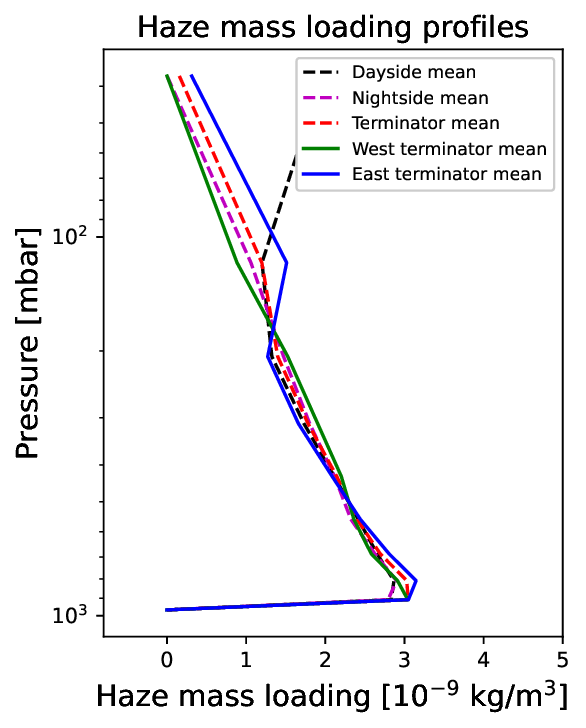}{0.33\textwidth}{d) Wolf 1061 c}}
\caption{Vertically integrated haze mass column for a) TRAP and b) WOLF and vertical haze mass profiles for c) TRAP and d) WOLF for regime 3. The rotation period shown is 6 days.}
\label{fig:reg3mmr}
\end{minipage}%
%\end{figure}

%\begin{figure}[ht!]
\begin{minipage}{0.99\textwidth}
\centering
\gridline{\fig{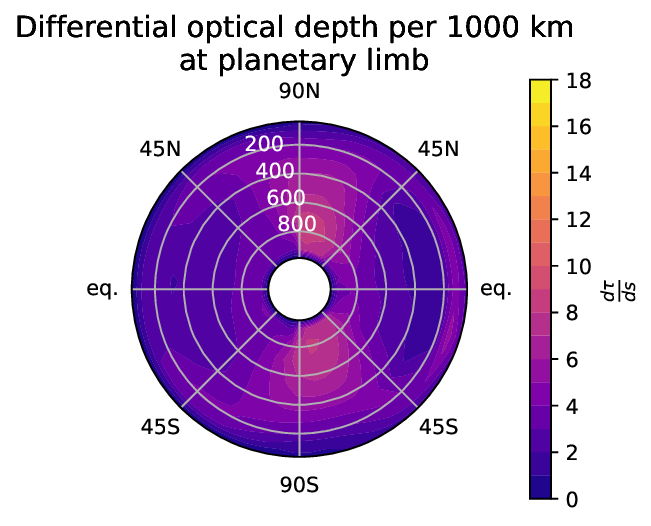}      {0.33\textwidth}{a) TRAPPIST-1 e}
          \fig{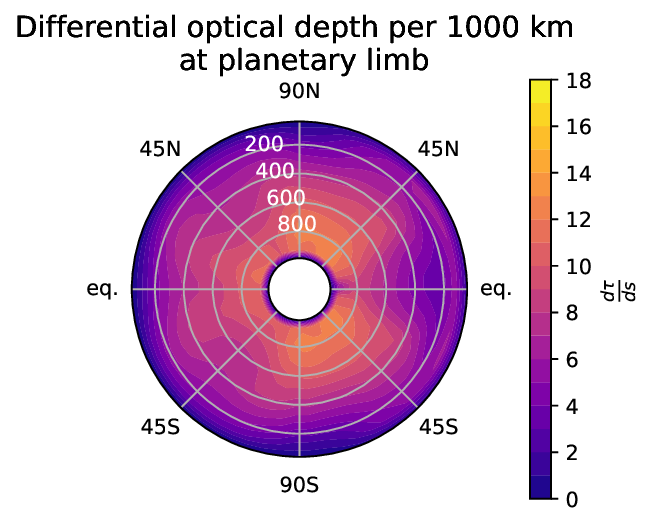}      {0.33\textwidth}{b) Wolf 1061 c}}
\caption{Differential optical depth per 1000 km in shortwave band 1 at the planetary limb for a) TRAP and b) WOLF for regime 3. The white text labels show pressure in mbar. The eastern (evening) terminator is shown on the right-hand side of each plot and the western (morning) terminator on the left-hand side of each plot to maintain consistency with other figures. The rotation period shown is 6 days.}
\label{fig:reg3tau}
\end{minipage}%
\end{figure}

%
For rotation periods of 3 (TRAP) or 4 (WOLF) to 12 days, both planets develop a circulation characterised by two rapid mid-latitude eastward jets. This double jet circulation regime is a common equilibrium state for simulations of TRAPPIST-1 e with its original rotation period of 6.1 days, predicted by 3 out of 4 models in the TRAPPIST-1 Habitable Atmosphere Intercomparison \citep{sergeev_trappist-1_2022}. As shown in Figure \ref{fig:reg3winds}, WOLF has substantially faster jets than TRAP, particularly near the model top. Also characteristic of this state are a pair of high-latitude Rossby gyres found near the eastern terminator, along with a less prominent matching pair at the western terminator. The gyres are key to the haze distribution in this regime. The vertical haze columns shown in Figure \ref{fig:reg3mmr} display several notable features: a much hazier WOLF atmosphere than TRAP, a reversed meridional haze gradient with a clearer equator and hazier poles compared to the banded regime, the build-up of a thick haze layer in the eastern Rossby gyres, and a small north-south asymmetry in haze mass. 

North-south asymmetry in simulations of tidally locked planets has been reported in other models, notably by \cite{braam_stratospheric_2023} in ozone columns, by \cite{landgren_shallow_2023} in the size of the Rossby gyres, and by \cite{noda_circulation_2017} in the surface temperature and other climate variables. This behaviour occurs in both sophisticated GCMs (the Met Office Unified Model in \cite{braam_stratospheric_2023}) and shallow-water models (the Shallow-Water Atmospheric Model in Python for Exoplanets in \cite{landgren_shallow_2023}). \cite{landgren_shallow_2023} hypothesise that a supercritical pitchfork bifurcation leads to north-south hemispheric asymmetry because at the convergence zone around the western terminator, where the eastward flow between the gyres encounters the westward dayside-to-nightside flow, even a small perturbation in the eastward winds leads to an uneven transfer of momentum across the equator to either the northern or southern hemisphere. The asymmetry of the gyres then causes asymmetries in the ozone column of \cite{braam_stratospheric_2023} and the surface temperature of \cite{noda_circulation_2017} because the gyres collect tracers and are associated with a surface temperature minimum.

The Rossby gyres are key to the high terminator differential optical depth in this regime, as shown in Figure \ref{fig:reg3tau}. \cite{braam_stratospheric_2023} describe how the overturning circulation of the troposphere and stratosphere on a slowly rotating tidally locked planet can transport tracer-enriched air from the dayside to the nightside and deposit it in the gyres. A similar pattern of tracer transport is reported in \cite{parmentier_3d_2013} (their Figure 6 for 0.5 $\mu$m particles) and \cite{steinrueck_3d_2021} (their Figure 7 for 1.0 $\mu$m particles). The large latitudinal extent of the gyres and their presence straddling the eastern terminator leads to the poor observing prospects of this simulated regime compared to the others.

\subsection{Single jet circulation regime}\label{subsec:reg4}
\begin{figure}
\centering
\gridline{\fig{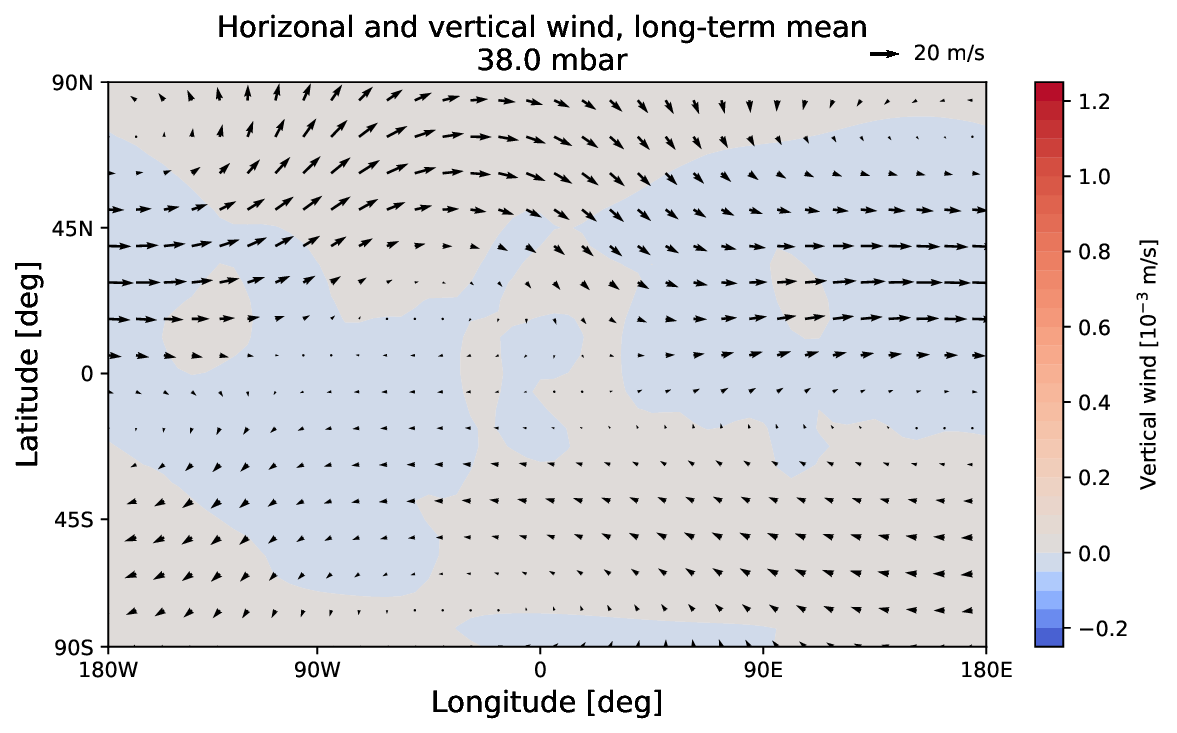}      {0.45\textwidth}{a) TRAPPIST-1 e, top of model }
          \fig{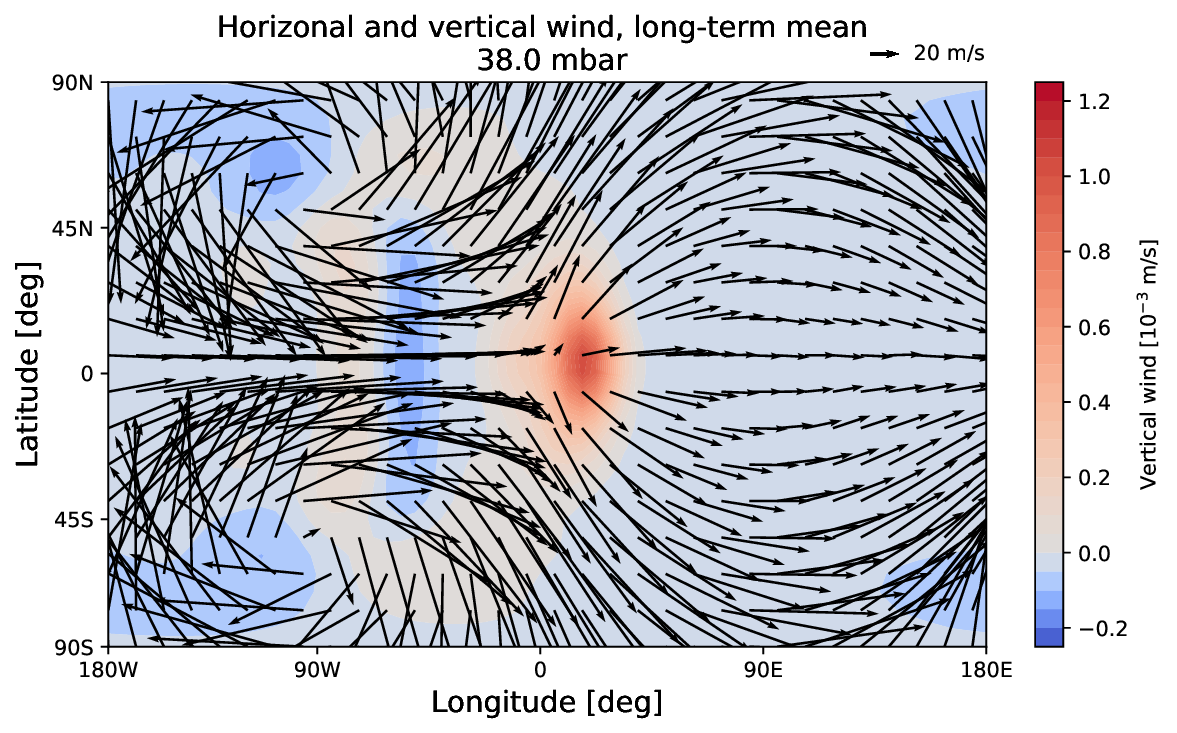}      {0.45\textwidth}{b) Wolf 1061 c, top of model }}
\gridline{\fig{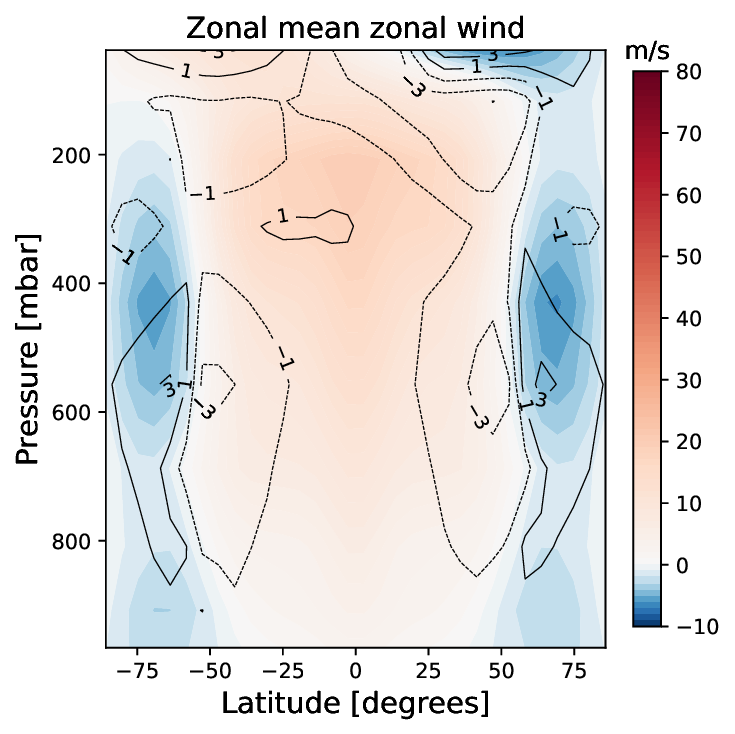}{0.33\textwidth}{c) TRAPPIST-1 e, zonal mean zonal wind}
          \fig{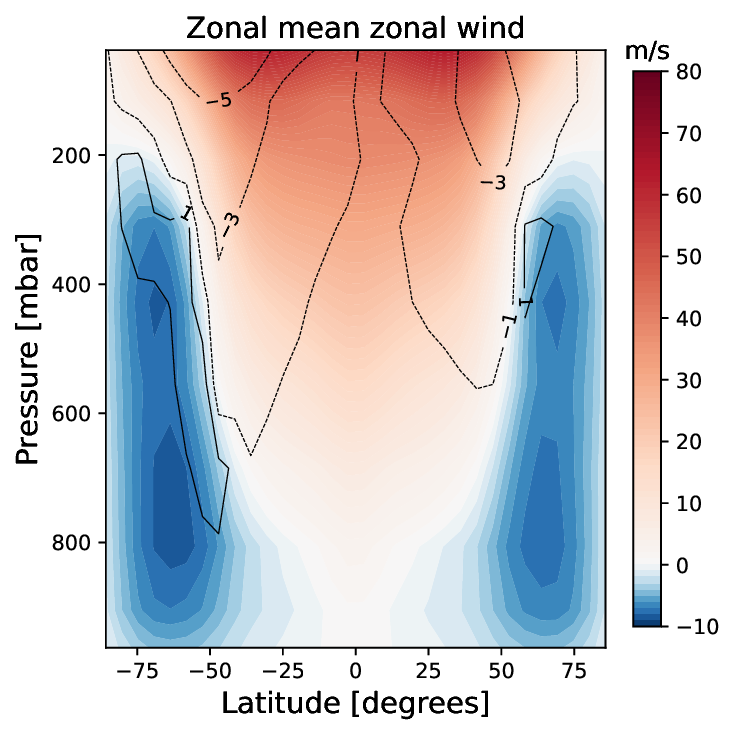}{0.33\textwidth}{d) Wolf 1061 c, zonal mean zonal wind}}
\caption{Top row: Horizontal and vertical winds in circulation regime 4 for a) TRAP and b) WOLF. Bottom row: Zonal mean zonal wind for c) TRAP and d) WOLF, with contour lines showing the difference in wind speed (absolute value) between the haze simulation and an identical simulation without haze (hazy minus control). Solid lines indicate positive values and dashed lines indicate negative. The rotation period shown is 18 days.}
\label{fig:reg4winds}
\end{figure}%
For periods of 13 to 30 days, ExoPlaSim simulates both TRAP and WOLF as planets with a single eastward jet centred at the equator. This circulation regime has been described in previous studies of rotation rate \cite{noda_circulation_2017, edson_atmospheric_2011} and in investigations of the equatorial superrotation on both gas and terrestrial planets \citep{showman_atmospheric_2013, showman_equatorial_2011, showman_matsuno-gill_2010}. Figure \ref{fig:reg4winds} c) and d) shows a broad jet covering most of the planet in both the TRAP and the WOLF cases, with higher winds speeds on WOLF, particularly near the model top, and a weaker westward flow near the poles. In the transition from the double jet to the single jet regime, the eastern Rossby gyres increase in size and migrate to the nightside of the planet. 

\begin{figure}
\begin{minipage}{0.99\textwidth}
\centering
\gridline{\fig{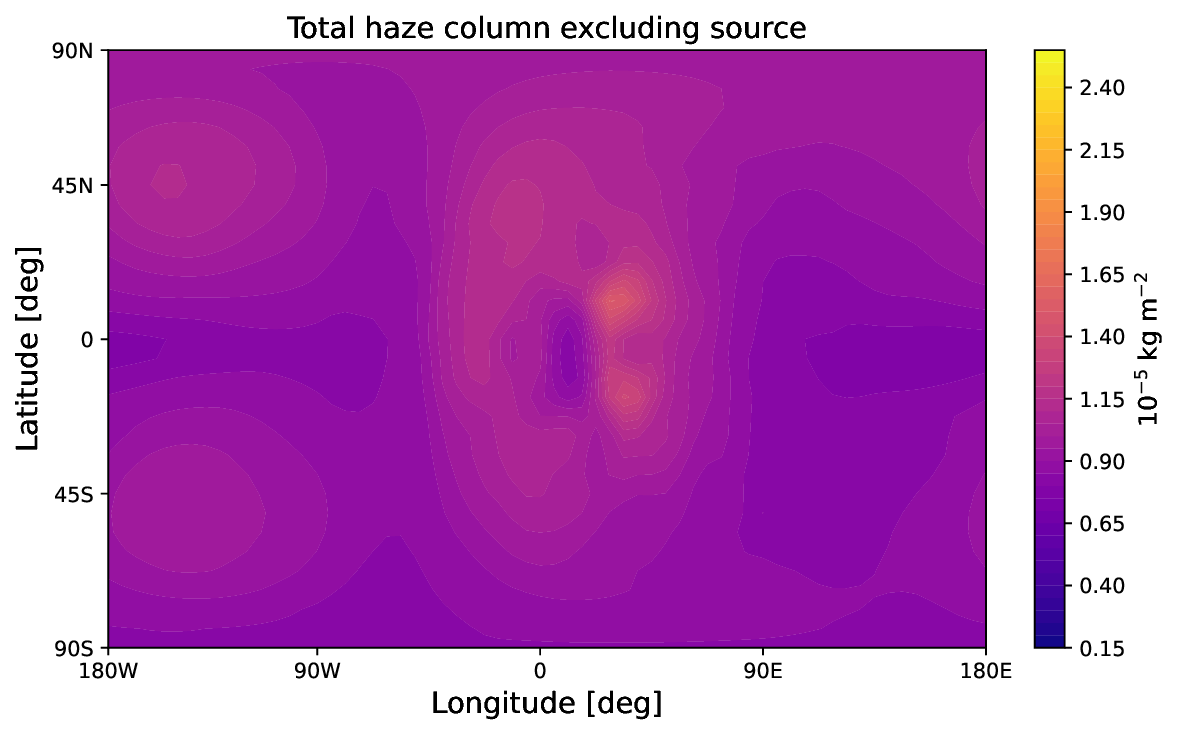}      {0.45\textwidth}{a) TRAPPIST-1 e}
          \fig{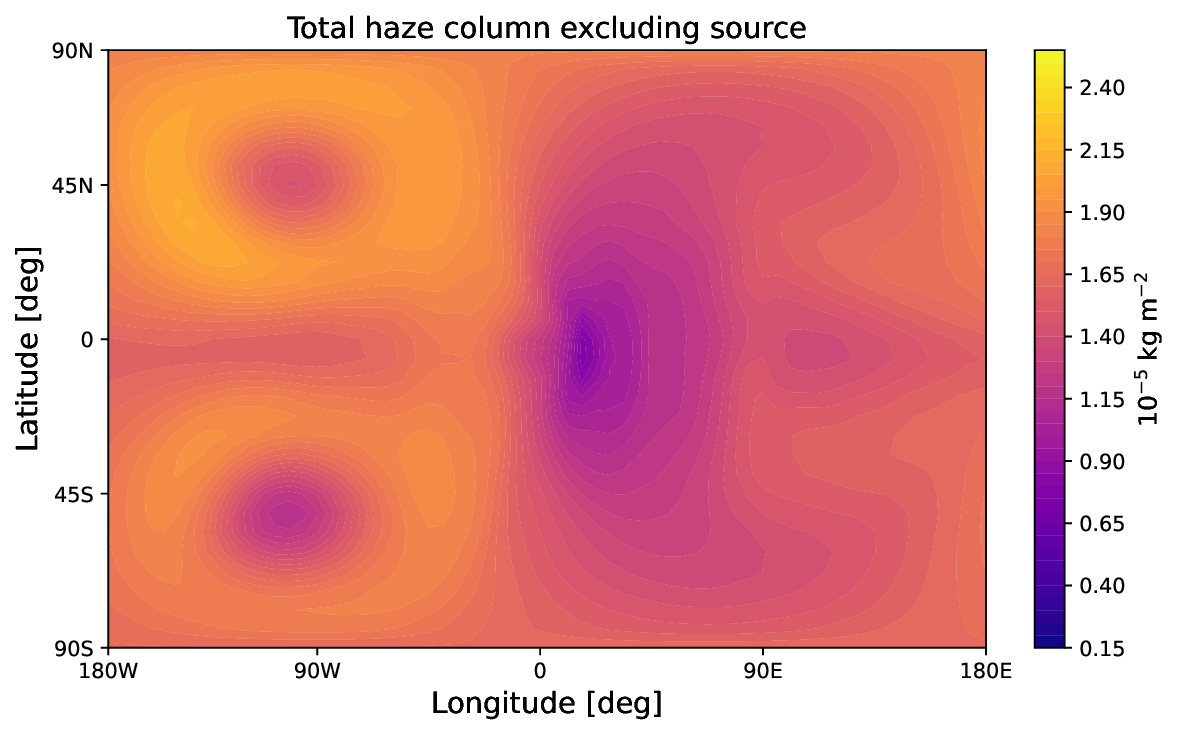}      {0.45\textwidth}{b) Wolf 1061 c}}
\gridline{\fig{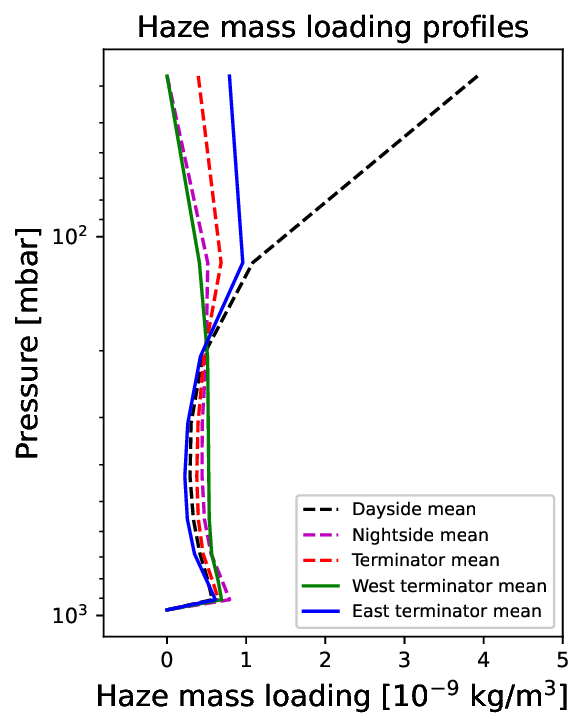}{0.33\textwidth}{c) TRAPPIST-1 e}
          \fig{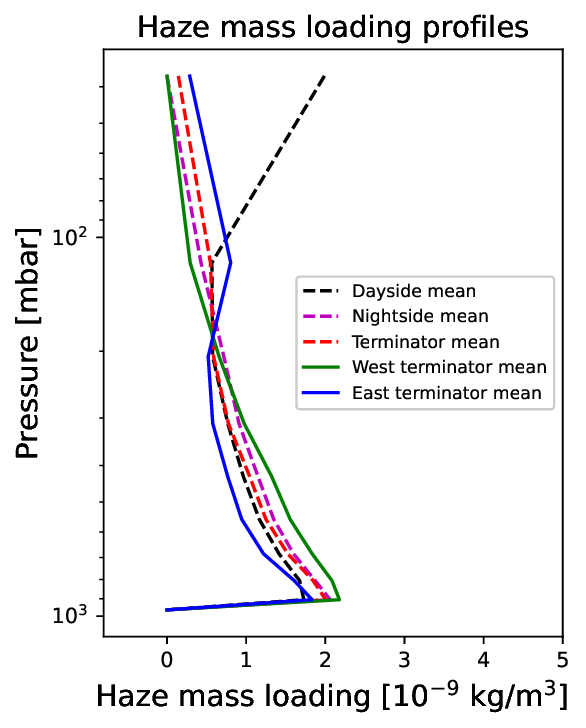}{0.33\textwidth}{d) Wolf 1061 c}}
\caption{Vertically integrated haze mass column for a) TRAP and b) WOLF and vertical haze mass profiles for c) TRAP and d) WOLF for regime 4. The rotation period shown is 18 days.}
\label{fig:reg4mmr}
\end{minipage}%
%\end{figure}

%\begin{figure}[ht!]
\begin{minipage}{0.99\textwidth}
\centering
\gridline{\fig{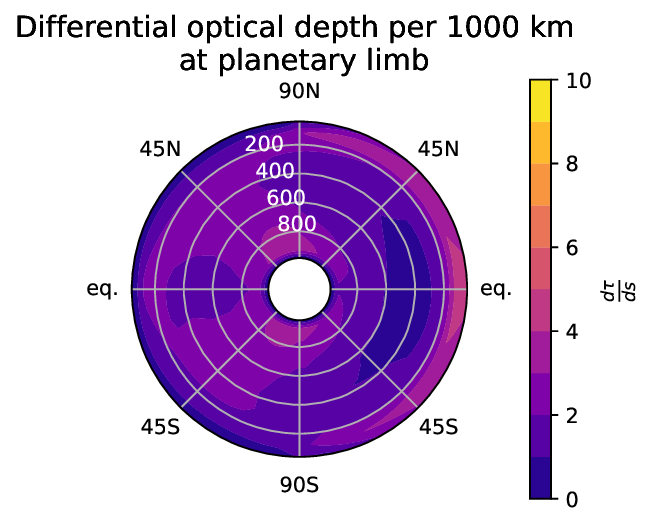}      {0.33\textwidth}{a) TRAPPIST-1 e}
          \fig{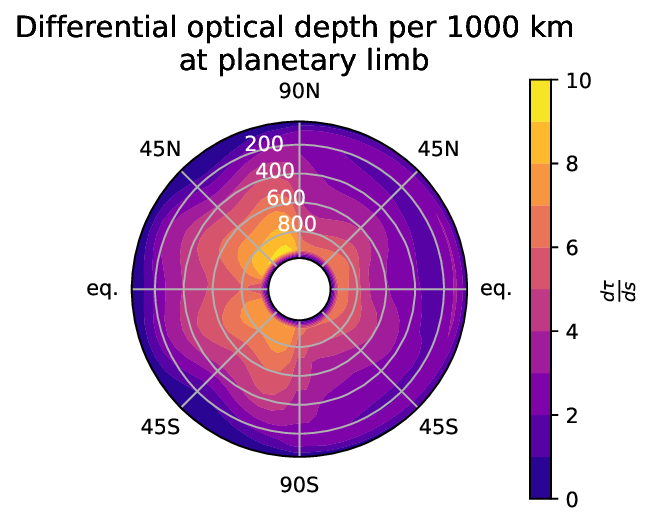}      {0.33\textwidth}{b) Wolf 1061 c}}
\caption{Differential optical depth per 1000 km in shortwave band 1 at the planetary limb for a) TRAP and b) WOLF for regime 4. The white text labels show pressure in mbar. The eastern (evening) terminator is shown on the right-hand side of each plot and the western (morning) terminator on the left-hand side of each plot to maintain consistency with other figures. The rotation period shown is 18 days.}
\label{fig:reg4tau}
\end{minipage}%
\end{figure}

The position of the gyres is once again key to the terminator differential optical depth shown in Figure \ref{fig:reg4tau}. If the gyres continue to collect haze particles but are now located on the nightside rather than the planetary limb, we might expect a less hazy terminator. This is the case for the TRAP simulation: the vertical haze column in Figure \ref{fig:reg4mmr} a) is highest in the substellar region and within the nightside gyres, and comparatively low elsewhere. In the WOLF case, however, (Fig. \ref{fig:reg4mmr} b)), the haze column is lowest in the substellar region and within the nightside gyres and highest in the region of high winds around the gyre centres. This results in a very high haze mass at the limb, this time at the western terminator. The vertical haze profiles shown in Fig. \ref{fig:reg4mmr} c) and d) reveal that TRAP has a high haze layer at altitudes above 200 mbar and less haze throughout the atmosphere below, while in the WOLF simulation haze mass increases steadily with atmospheric pressure. This vertical structure is consistent with \cite{braam_stratospheric_2023}'s description of tracer transport by an overturning dayside-nightside circulation from a high source region on the dayside to the nightside gyres. As the TRAP case develops a high haze layer, the overturning circulation can transport haze from this layer into the gyre centres. In contrast, haze settles more rapidly in the WOLF simulation and experiences the edges of the gyres as a barrier when horizontally transported from dayside to nightside.

\clearpage
\subsection{Parameter space overview}\label{subsec:spaceoverview}
\subsubsection{Observing region}\label{subsubsec:observing}

\begin{figure}
\gridline{\fig{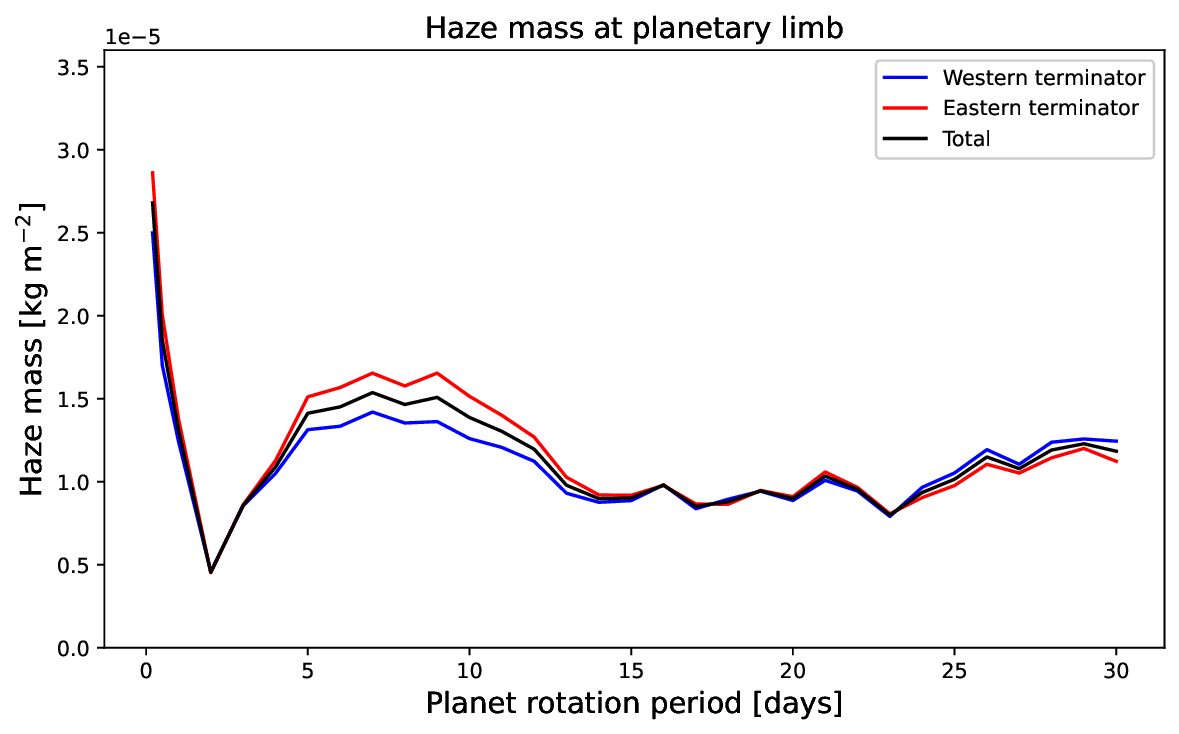}      {0.45\textwidth}{a) TRAPPIST-1 e}
          \fig{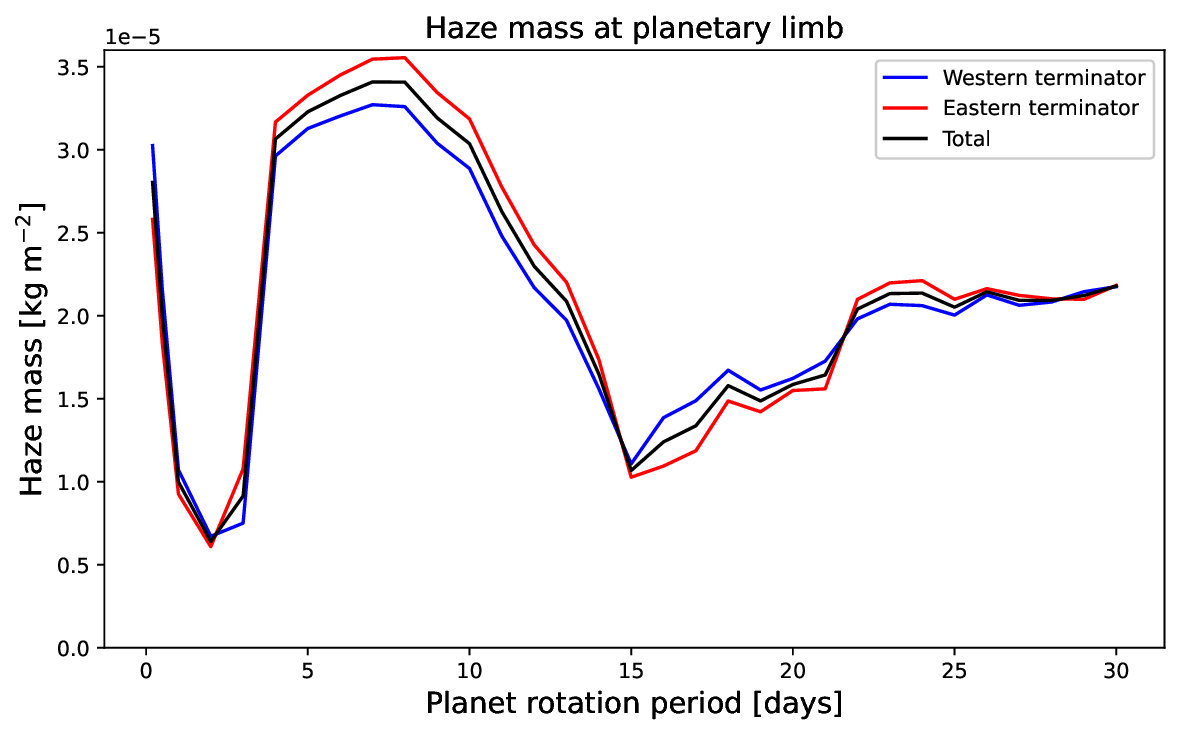}      {0.45\textwidth}{b) Wolf 1061 c}}
\caption{Haze mass at the planetary limb in kg/m$^2$ as a function of rotation rate for a) TRAP and b) WOLF. Haze mass loading is integrated downwards as in the vertically integrated haze column and then additionally along the limb to account for differences in spacing between latitudes. The total haze mass at the limb is then divided by the surface area of the limb to further account for the difference in size between the two planets, resulting in a unit of kg/m$^2$}.
\label{fig:limbmass}
\end{figure}

The four regimes described above are representative of our chosen parameter space. We now give an overview of the full parameter space as a function of rotation rate. Figure \ref{fig:limbmass} displays the haze mass per meter squared at the planetary limb as a function of planetary rotation period. The four circulation regimes are clearly visible. Haze mass, initially high at short periods in the banded regime, drops steeply with rotation rate to the low-haze valley at 2 (TRAP) or 3 (WOLF) days. It then rises again in the double jet regime, but there are clear differences between the two planets: the TRAP case takes the rough form of a parabola, peaking at around 8 days in the middle of the double jet regime. The haze mass remains lower than in the banded regime throughout. In contrast, haze mass increases sharply in the WOLF case to a maximum at 5 days. In the WOLF simulations, the double jet regime is even hazier than the banded regime. In the single jet regime, both planets see a gradual increase with rotation period, but the total haze mass remains substantially lower than in the banded and double jet regimes. Fig. \ref{fig:limbmass} also reveals systematic patterns in terminator asymmetry. In the double jet regime, both TRAP and WOLF have more haze mass at the eastern terminator, while in the single jet regime, the WOLF simulations now have a hazier western terminator up to a 22-day rotation period and the TRAP simulations show little asymmetry. This difference is due to the trapping of haze particles in the nightside gyres in the TRAP simulations but not in the WOLF simulations in the single jet regime. 

\begin{figure}
\gridline{\fig{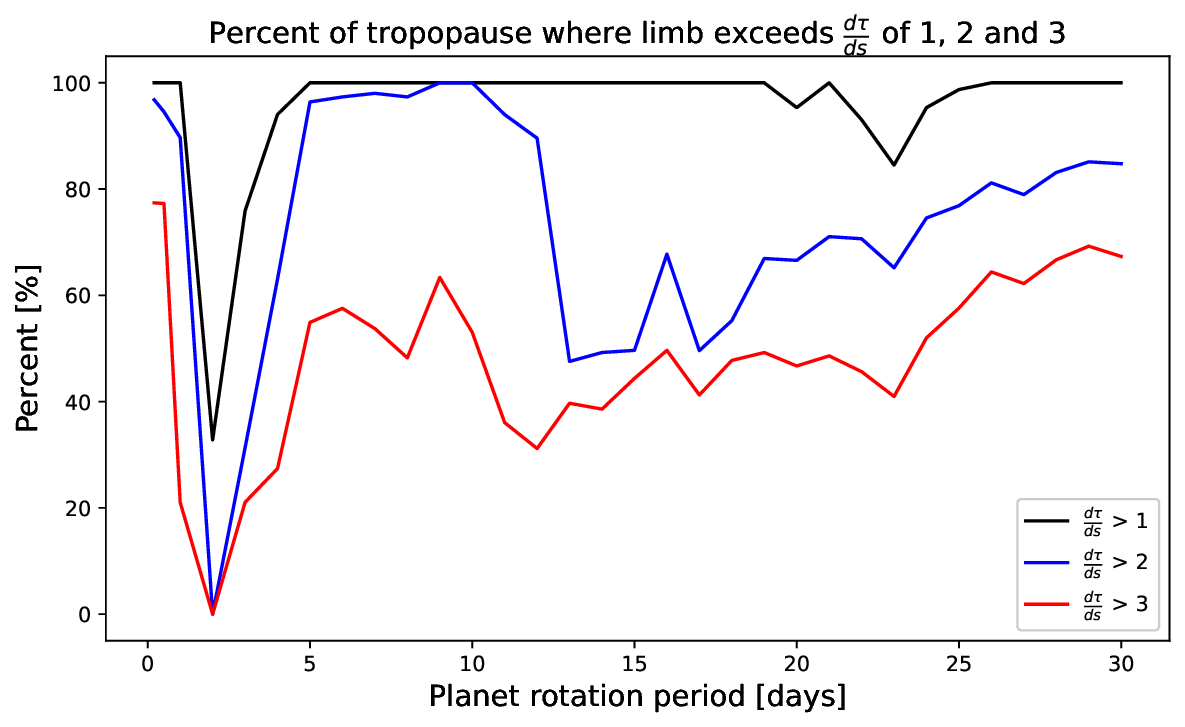}      {0.45\textwidth}{a) TRAPPIST-1 e}
          \fig{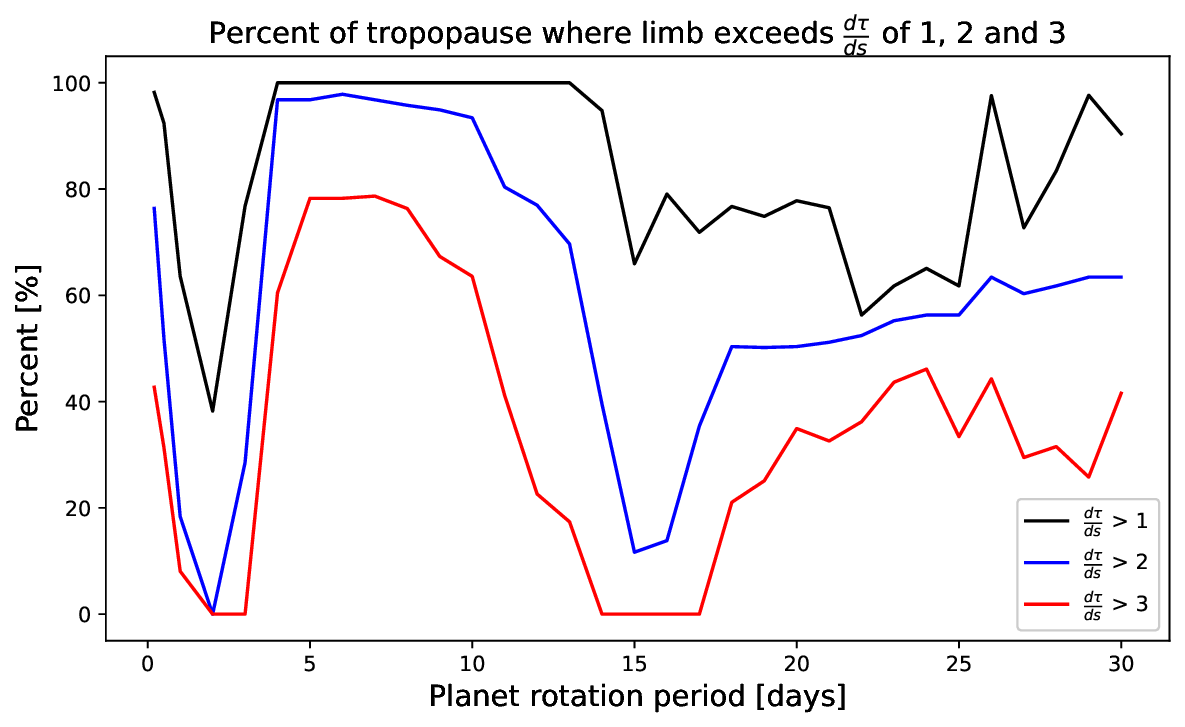}      {0.45\textwidth}{b) Wolf 1061 c}}
\gridline{\fig{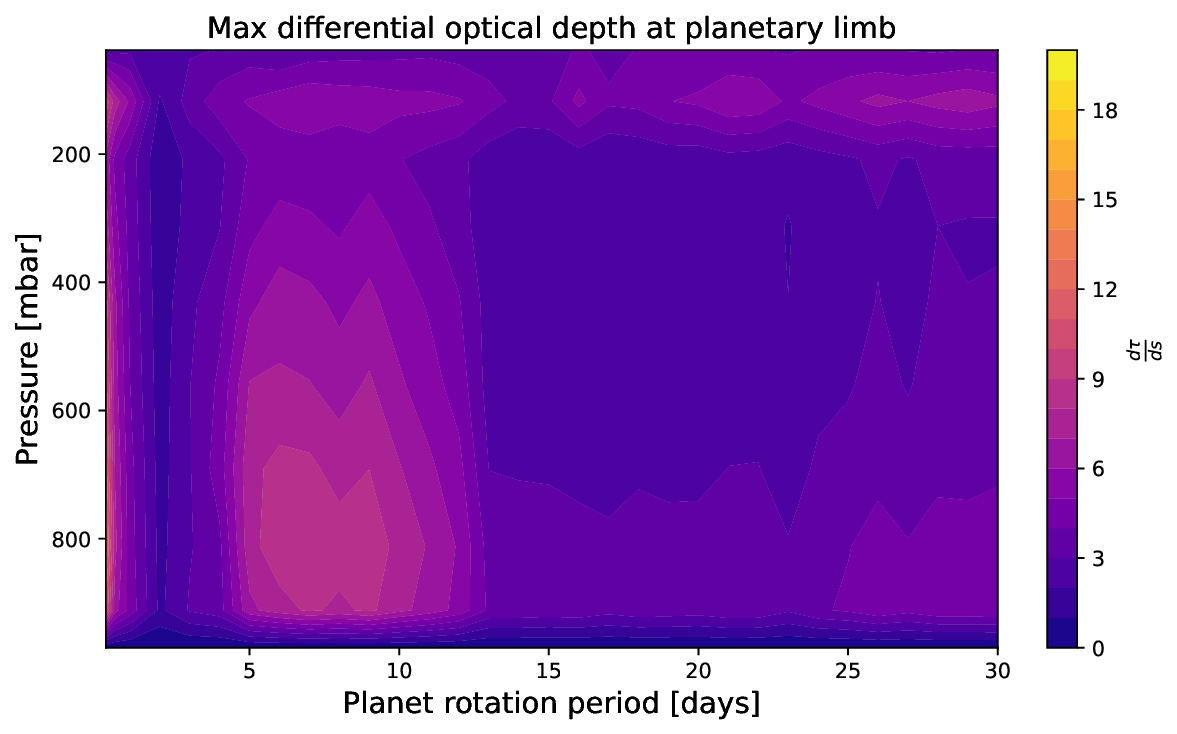}      {0.45\textwidth}{c) TRAPPIST-1 e}
          \fig{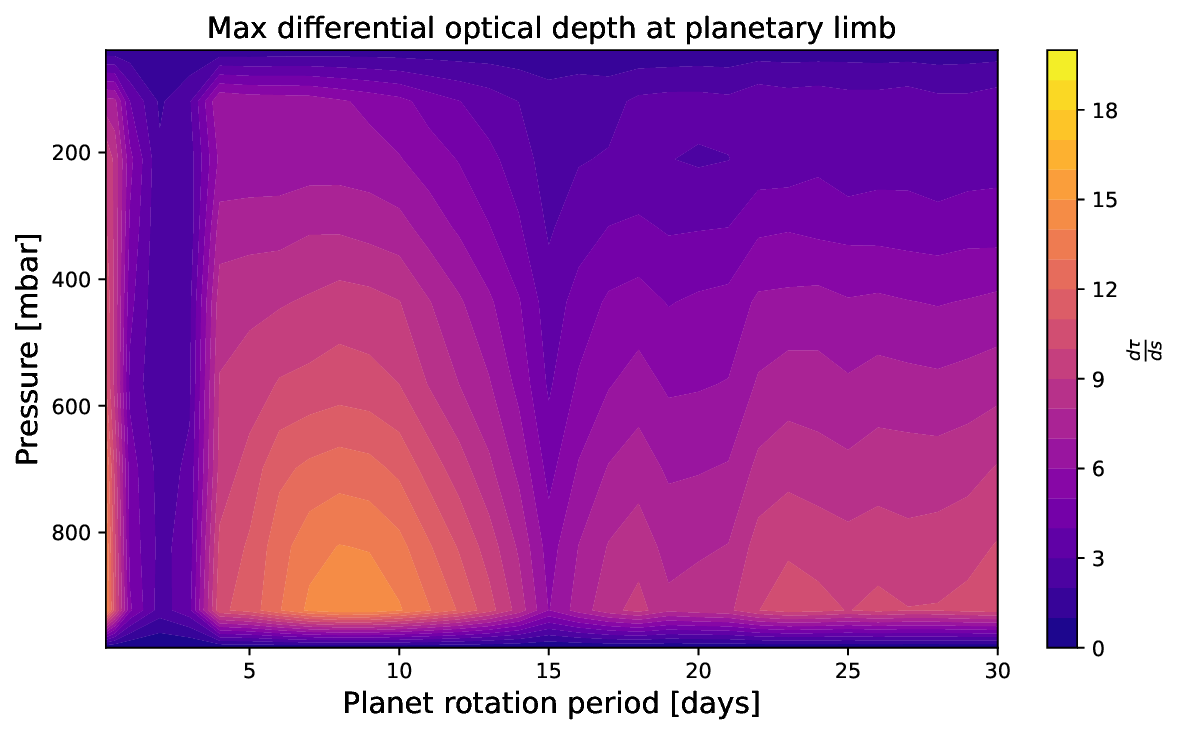}      {0.45\textwidth}{d) Wolf 1061 c}}
\caption{Top row: Percentage of tropopause (top model level) with a differential optical depth for shortwave band 1 greater than 1, 2, and 3 at the planetary limb for a) TRAP and b) WOLF. Bottom row: Maximum differential optical depth for shortwave band 1 at the planetary limb as a function of atmospheric pressure and rotation rate for c) TRAP and d) WOLF. The differential optical depth is calculated for a fixed path length of 1000 km as in Fig. \ref{fig:reg1tau}, \ref{fig:reg2tau}, \ref{fig:reg3tau}, and \ref{fig:reg4tau}. As the value is sensitive to the choice of path length, these figures express the relative haziness of the atmosphere in different circulation regimes and at different altitudes, rather than a quantitative prediction of optical thickness.}
\label{fig:limbtau}
\end{figure}

While Fig. \ref{fig:limbmass} could suggest that slowly rotating planets are more favourable observing targets because they generally have lower differential optical depths at the limb, it does not account for the height at which haze particles accumulate. The height of the haze layer is as important as the particle abundance, as instruments cannot probe below the haze layer if the atmosphere exceeds a certain threshold of optical depth. To quantify the impact of vertical structure on potential observations, we show in Figure \ref{fig:limbtau} the percentage of the top model level (the tropopause) which exceeds differential optical depths of 1, 2, and 3 in shortwave band 1. Fig. \ref{fig:limbtau} a) and b) mirror Fig. \ref{fig:limbmass} with respect to the banded regime, haze valley, and double jet regime: the terminators are largely impenetrable in the banded and double jet regimes and anomalously clear in the transitional regime between them. However, in the single jet regime, differential optical depths are higher in more of the limb for TRAP than for WOLF, even though TRAP has less haze mass overall. 

The reason for this is revealed in Fig. \ref{fig:limbtau} c) and d), which show the maximum differential optical depth at each atmospheric pressure level as a function of rotation period. The hazy banded regime, clear transitional regime, and hazy double jet regime are visible and consistent with Fig. \ref{fig:limbmass} and Fig. \ref{fig:limbtau} a) and b), but the difference in vertical structure in the single jet regime is now apparent. Haze particles continue to build up in the deeper atmosphere in the WOLF simulations as rotation period increases, but do not form a layer at the top of the model as in the TRAP simulations.

\subsubsection{Climate impact}\label{subsubsec:climpact}
\begin{figure}
\gridline{\fig{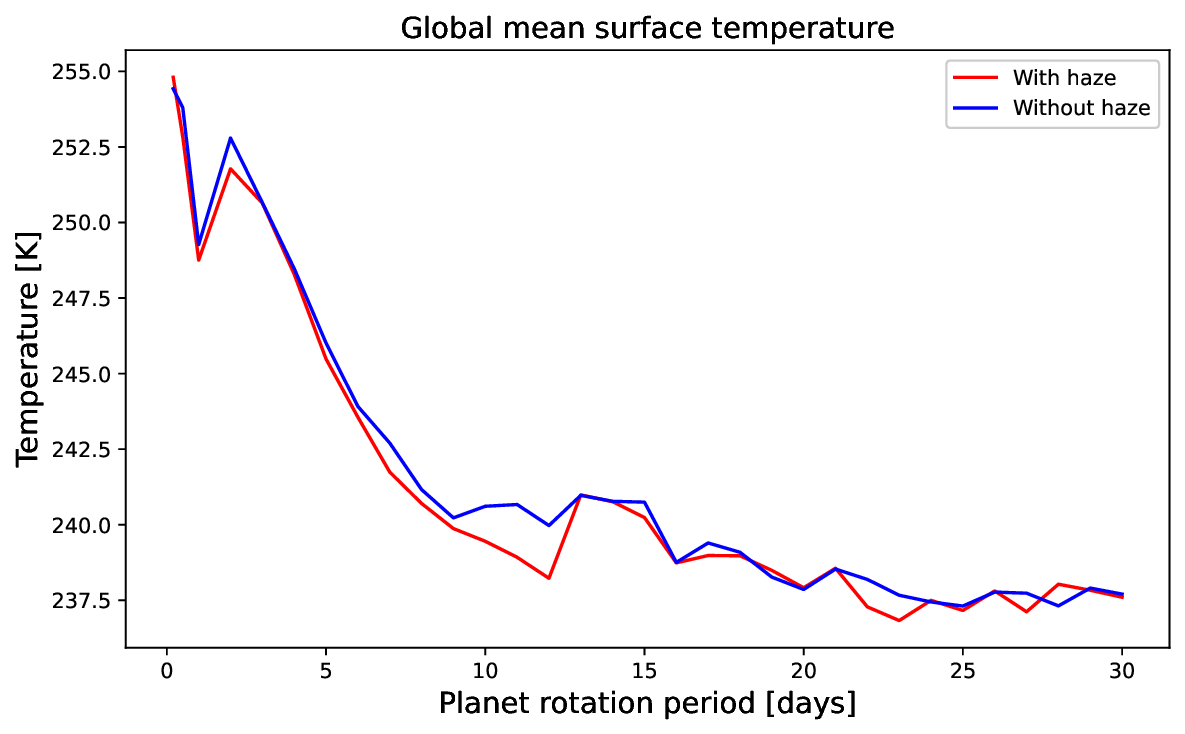}      {0.45\textwidth}{a) TRAPPIST-1 e}
          \fig{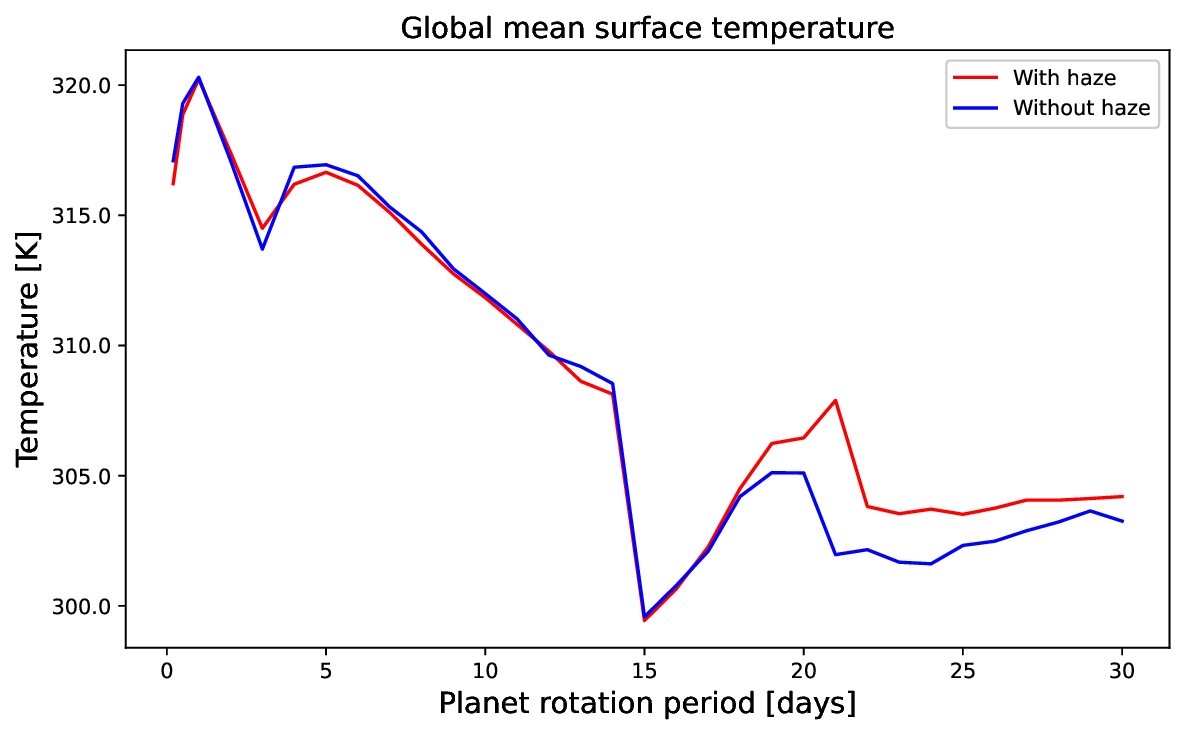}      {0.45\textwidth}{b) Wolf 1061 c}}
\gridline{\fig{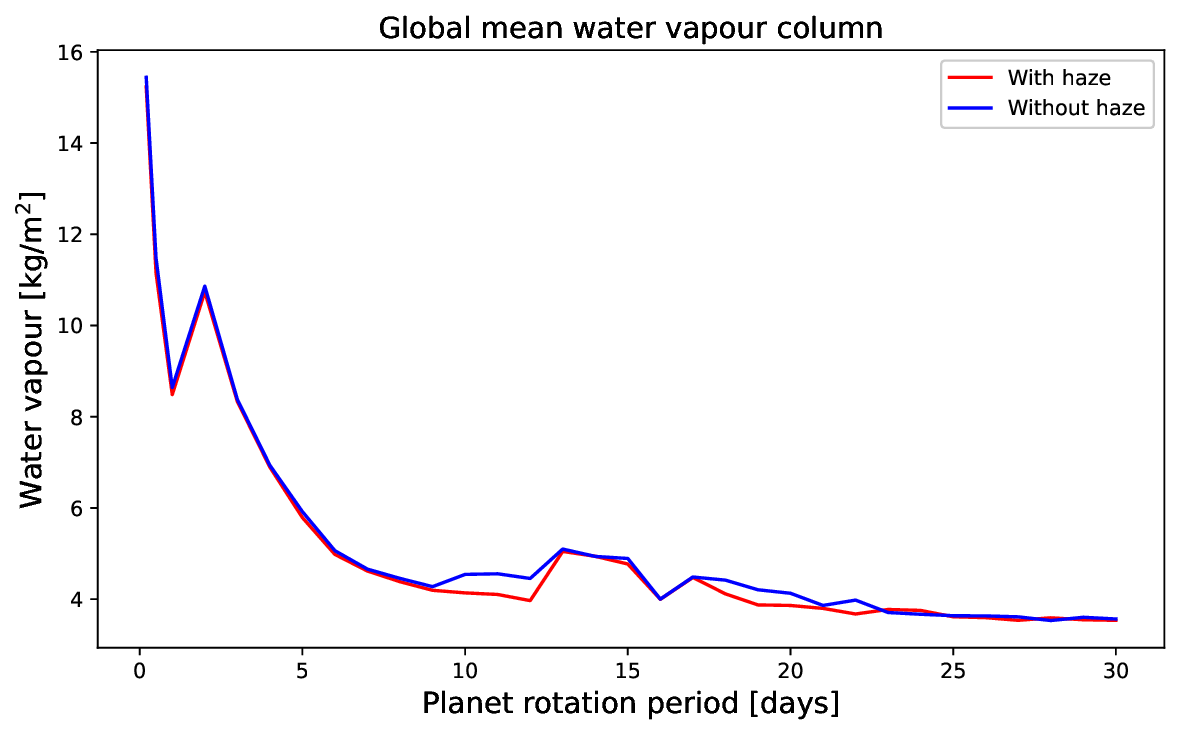}      {0.45\textwidth}{c) TRAPPIST-1 e}
          \fig{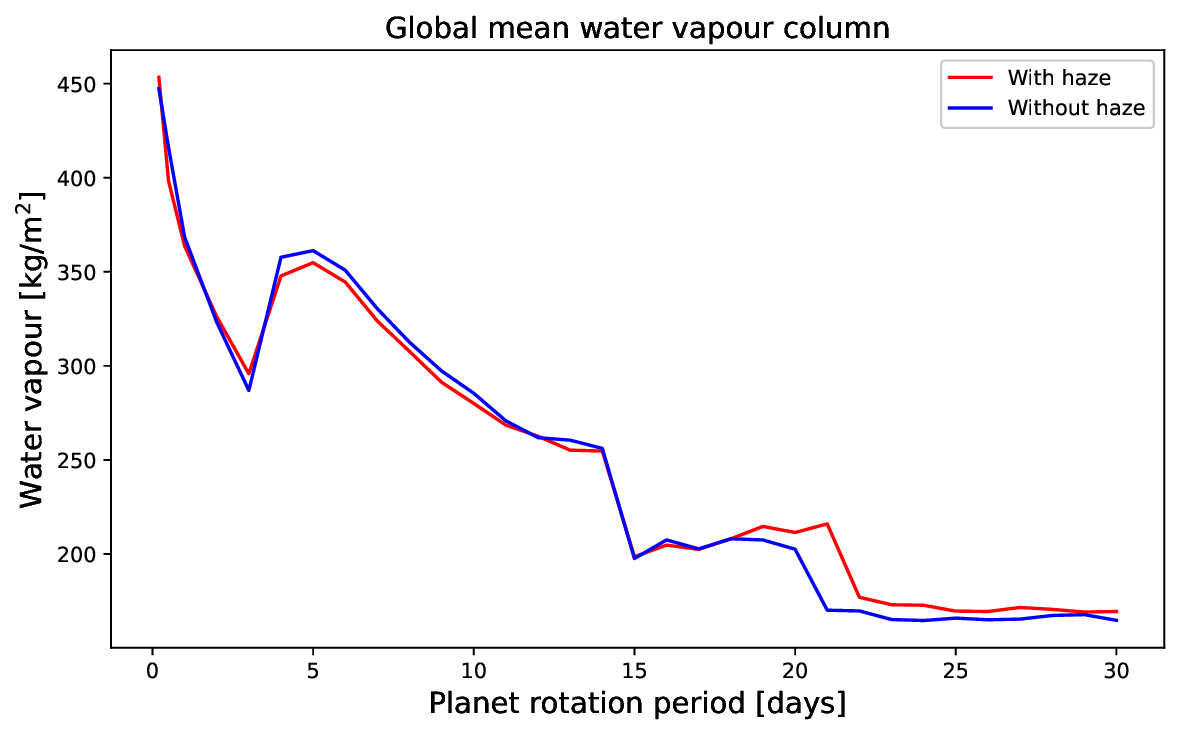}      {0.45\textwidth}{d) Wolf 1061 c}}
\caption{Effect of radiatively active organic haze particles on the global mean surface temperature for a) TRAP and b) WOLF and water vapour column for c) TRAP and d) WOLF as a function of rotation period.}
\label{fig:climpact}
\end{figure}

To examine the impact of radiatively active organic haze particles on the planetary climate, we show the global mean surface temperature and global mean water vapour column in Figure \ref{fig:climpact} for both our simulations with radiative effects of haze and the matching set of control simulations without haze. For most of the parameter space, the haze has little impact on the climate and effectively acts as a passive tracer. Table \ref{tab:mieqs} shows that most of the extinction is caused by scattering rather than absorption, and that the scattering is mostly in the forward direction, as the backscattering efficiencies are small. It may be that the particles scatter light forwards through the atmosphere and towards the surface, leading to similar instellation at the surface as a clear atmosphere and therefore a similar surface temperature and water vapour column. For the TRAP case, some simulations in the double jet regime show a small anti-greenhouse effect of no more than 1-2 K, while in the WOLF case, the slowest rotators in the single jet regime show warming of up to 5 K due to the haze. The contours in Figures \ref{fig:reg1winds}, \ref{fig:reg2winds}, \ref{fig:reg3winds}, and \ref{fig:reg4winds} c) and d) in turn show the radiative feedback of the haze on the general circulation, with positive values corresponding to regions where the hazy simulation predicts higher wind speeds than the control simulation. In general, the differences in wind speed increase with altitude. In the banded regime, the westward equatorial jet slows while the eastward mid-latitude jets accelerate. In the double jet regime, the TRAP case exhibits acceleration of the southern mid-latitude jet and slowing of the northern mid-latitude jet compared to the control, which we interpret as a perturbation in the north-south asymmetry rather than a shift in the circulation regime itself. In the WOLF case, the wind speeds are nearly identical, with very slight acceleration (+1 m/s) of both mid-latitude jets. In the single jet regime, the single equatorial jet slows for both planets. The transitional regime in the TRAP case has slightly faster mid-latitude jets than the control simulation, while the WOLF cases are again nearly identical. The magnitude of the difference is small in all simulations, with 1-3 m/s typical for most of the atmosphere. 

\subsection{Sensitivity to particle size and density}\label{subsec:sensitivity}
We performed a preliminary parameter space study of particle size and density to test the gravitational settling scheme, as well as the sensitivity of the haze distribution to these inputs. The particle sizes and densities tested are described in Section \ref{sec:methods}. As our results broadly reproduce the findings of \cite{steinrueck_3d_2021} and \cite{parmentier_3d_2013}, which contain detailed analyses of the settling behavior, we only summarise them here. The haze distribution falls into either a ``small particle'' regime, in which the particles are not affected by gravitational settling and the equilibrium mass mixing ratios are insensitive to size and density, or a ``large particle'' regime in which haze particles begin to settle under gravity and collect in the lower atmosphere. In the large particle regime, total haze mass decreases with particle size and density, as particles settle out of the atmosphere faster. The transition between the regimes occurs at around 100 nm, placing our chosen 500 nm particles well into the large particle regime. We also found that a particle size of 500 nm resulted in the largest extinction efficiencies for our given haze refractive indices and stellar spectra. We therefore expect that our simulated particle size will have the most substantial climate impact out of the large particles in our tested range. Based on these tests, we believe the results presented above are robust to particle size and density for the given initial haze source and our two planets. However, our findings -- including that the particles act as passive tracers -- do not necessarily generalise to other planets or stars, whose spectra will result in different Mie efficiencies, or to a higher haze source than used in our simulations.

\section{Discussion}\label{sec:discussion}

\subsection{Circulation regimes}\label{subsec:circreg}
The progression of circulation regimes with rotation period parallels the study of moist atmospheric circulation as a function of rotation period conducted by \cite{noda_circulation_2017}, \cite{haqq-misra_demarcating_2018}, and \cite{edson_atmospheric_2011}, and to a lesser extent the dry circulation regimes modelled by \cite{carone_connecting_2015}, and \cite{merlis_atmospheric_2010}. There is good agreement between models when it comes to short-period planets: \cite{noda_circulation_2017}'s type-IV circulation, \cite{haqq-misra_demarcating_2018}'s rapid rotators, and tidally locked simulations with a 1-day period in \cite{carone_connecting_2015}, \cite{edson_atmospheric_2011}, and \cite{merlis_atmospheric_2010} all resemble our banded regime.

While both the single and double jet states appear in simulations of tidally locked planets, not all studies distinguish between them and, even if both are identified, the transitions between them occur at different rotation periods. \cite{noda_circulation_2017} describe their 
type-II circulation, applicable to their simulations with rotation periods from 6.7-20 days, as characterised by a stationary Rossby wave on a broad equatorial westerly jet. This description corresponds to our single jet regime (13 to 30 day rotation periods). The authors note that as the simulated planet spins faster, the equatorial jet slows as transient mid-latitude disturbances begin to transport eastward momentum from the equator toward the poles, a mechanism indicating a shift to a mid-latitude jet state. In a study of a climate bistability between the double jet and single jet circulation regimes for sophisticated GCM simulations of TRAPPIST-1 e (rotation period of 6.1 days), \cite{sergeev_bistability_2022} investigates a phenomenon in which some simulation initial conditions result in a transfer of angular momentum towards the equator (single jet regime) and others towards the poles (double jet regime). Their Figure 7 depicts the single and double jet regimes as nearly identical to those in ExoPlaSim in terms of Rossby gyre location and jet structure. \cite{haqq-misra_demarcating_2018} categorise planets in both of these states as ``Rhines rotators'', a regime characterised by a combination of a thermally direct overturning circulation from dayside to nightside and the presence of zonal jets. Their Figure 9 shows both a single jet structure (the 3300 K case) and a double jet structure (the 3000 K case). However, as their study varies stellar type, stellar constant, and rotation period in tandem to represent physically consistent tidally locked planets, there are several confounding factors compared to our parameter space of rotation period only. \cite{carone_connecting_2015} in turn define circulation regimes based on the ratio of the tropical and extratropical Rossby radius of deformation to the planet's radius, i.e., based on whether tropical or extratropical Rossby waves can ``fit'' on the planet. Their tropical Rossby wave-dominated regime matches our single jet regime and their extratropical Rossby wave-dominated regime matches our double jet regime  (schematically shown in their Figure 2) with respect to the position of the Rossby gyres. However, they find that slower rotators tend towards two mid-latitude jets and faster rotators towards a single fast equatorial jet. \cite{edson_atmospheric_2011} in contrast find mid-latitude jets at rotation periods of 3 and 4 days, as well as a structure that could be a single jet or transitional between double and single jets at 5 days (their Figure 4). As their parameter space only covers periods of 1, 3, 4, 5, and 100 days, it is not possible to make further comparisons.

If the single and double jet states can indeed be viewed as variations of a single regime (the Rhines rotators of \cite{haqq-misra_demarcating_2018} or type-II circulation of \cite{noda_circulation_2017}) distinguished by whether angular momentum is transferred either towards the equator or towards the poles \citep{sergeev_bistability_2022}, different studies may have inconsistent results for the same rotation period if other factors, such as choice of temperature forcing or stellar constant, dry vs. moist atmosphere, or the presence clouds, are not kept constant or if implementations of model physics differ. As our work shows, however, distinguishing between these states is crucial to explaining the haze distribution, as the location of the Rossby gyres is central to how hazy the terminators become. Further work on the emergence of the single and double jets in the vein of \cite{sergeev_bistability_2022}, particularly in more idealised models that enable better investigation of underlying mechanisms, is necessary to clarify the relationship between these states and the various factors that influence them. Conversely, if it becomes possible to diagnose haze distribution through, for example, asymmetry in the haze layer altitude between the leading and trailing limb of a transmission spectrum, this could be evidence for a particular circulation regime.

Finally, with respect to the low-haze transitional regime between the banded and double jet regimes, we can compare our results to \cite{noda_circulation_2017} and \cite{carone_connecting_2015}. Although both studies simulated planets with 2-day rotation periods, neither display the results of this experiment. \cite{noda_circulation_2017} identify a type-III circulation regime with periods from 1.25 to 2 days, characterised by a circulation similar to that of type-II (broad eastward jet with stationary Rossby waves) but with significant north-south asymmetry in values such as surface temperature and geopotential height, as well as long-term periodic variability in this asymmetry. \cite{carone_connecting_2015} show results from their 3-day period simulation, which is characterised by the presence of both tropical and extra-tropical Rossby waves. As neither these results nor ours form a consistent picture, clearly further theoretical work is needed in understanding planets with rotation periods of approximately 1-3 days. Tidally locked planets with rotation periods this short are unlikely to be temperate, but may still form photochemical hazes.

\subsection{Distribution and effects of haze}\label{subsec:hazediscussion}
We can compare our results to the passive haze tracer studies of \cite{steinrueck_3d_2021} and \cite{parmentier_3d_2013} and the radiatively active haze tracer studies of \cite{steinrueck_photochemical_2023} and \cite{mak_3d_2023}. \cite{steinrueck_3d_2021} and \cite{parmentier_3d_2013} investigated gravitational settling and vertical mixing of passive haze tracers with varying radii in simulations of hot Jupiters HD 189733 b and HD 209458 b, respectively. As in our Section \ref{subsec:sensitivity}, in both studies haze particles fall into two regimes based on size: particles with a radius larger than the mean free path of the background gas fall with a terminal velocity independent of the pressure, while particles smaller than the mean free path have terminal velocities inversely proportional to the pressure. If all else is held equal, an N$_2$ atmosphere should have a larger mean free path than an H$_2$ atmosphere, shifting the transition from small to large particle regime towards larger particles. However, other factors such as gravity (21.93 m/s$^2$ in \cite{steinrueck_3d_2021} and 9.81 m/s$^2$ in \cite{parmentier_3d_2013}) and viscosity (dependent on local temperature, which is much higher in both hot Jupiter studies) also come into play. \cite{steinrueck_3d_2021} place the transition at 10-30 nm, compared to our 100 nm. In the small particle regime, \cite{steinrueck_3d_2021} also report particle trapping in nightside vortices which extend across the morning terminator, while in the large particle regime, the vortices are depleted of hazes.

In recent work, \cite{steinrueck_photochemical_2023} and \cite{mak_3d_2023} performed radiatively active haze tracer simulations of hot Jupiter HD 189733 b and the Archaean Earth, respectively, both using optical constants for organic hazes analogous to Titan's tholins reported in \cite{khare_optical_1984}. \cite{steinrueck_photochemical_2023} found that Titan-like hazes increase the net heating rate across a broad range of atmospheric pressures, significantly influencing the atmospheric circulation and therefore changing the haze distribution compared to the passive tracer simulations. In contrast,  \cite{mak_3d_2023} found that for the Archean Earth only a thin haze layer leads to atmospheric heating, while thicker haze layers (6 out of their 7 simulations) lead to atmospheric cooling. We note that although both the optical constants and the stellar spectra used in these studies differ from our own, \cite{mak_3d_2023}'s simulated atmosphere is similar to ours and indeed also based on the chemical compositions used in \cite{arney_pale_2017} (nitrogen atmosphere with 0.2 ratio of CH$_4$ to CO$_2$). 

While \cite{braam_stratospheric_2023} focus on ozone distribution, the mechanism they identify for transport of tracers from a dayside photochemical source to the nightside Rossby gyres is relevant to our results. This study shows that the overturning circulation on a tidally locked planet can advect tracers from the dayside in the upper atmosphere and deposit them in the areas of low geopotential height within the nightside gyres. The analysis primarily discusses the stratosphere because this is where ozone is produced, but the same mechanism could explain why haze particles collect in the nightside gyres in the TRAP simulations which form a high dayside haze layer, but not in the WOLF simulations which are lacking such a high layer.

As this discussion shows, drawing broad conclusions about the distribution and climate impact of photochemical hazes is difficult. This is not only because of the limited amount of literature available, but also because of the sensitivities that literature reveals. Haze particles may collect inside Rossby gyres or remain circulating outside them, depending on whether the gyres are fed by a high altitude haze layer. Haze may either warm or cool the planet's surface and atmosphere, depending on its optical properties and the thickness of the haze layer. Haze particles may affect the general circulation significantly if they absorb light and change the temperature structure of an atmosphere, or they may act as passive tracers. Understanding which of these effects is in play for a particular exoplanet cannot be predicted in advance, but requires observational data to guide modelling efforts.

\subsection{Implications for observations and future work}\label{subsec:implications}
In our study, we varied the rotation rate of each of our simulated planets while keeping all other parameters constant. This allows us to attribute changes across the parameter space solely to rotation rate and circulation regime, without having to consider the confounding effects of different haze production rates, stellar spectra, stellar constants, atmospheric compositions, and numerous other, often unconstrained factors. On the other hand, our simulations may not all correspond to real atmospheric states, as we do not co-vary rotation period, stellar constant, stellar type, and haze production rate, which are or could be linked for tidally locked planets. To better relate our simulations to the known exoplanet population, we identify below a number of temperate tidally locked rocky planets whose rotation periods place them within our parameter space.

The double jet TRAP simulation presented in Section \ref{subsec:reg3} is physically self-consistent, as it takes the rotation period (6 days), stellar spectrum, and stellar constant, as well as planet radius and mass from the real planet TRAPPIST-1 e. TRAPPIST-1 e is one of the most frequently modelled exoplanets and serves as a benchmark for GCM studies of rocky tidally locked worlds. Moist simulations of the planet as part of the Trappist-1 Habitable Atmosphere Intercomparison (THAI) predicted a double jet circulation regime in three out of four cases \citep{sergeev_trappist-1_2022}. The TRAPPIST-1 system has already been observed multiple times \citep{lim_atmospheric_2023, lincowski_potential_2023, zieba_no_2023, greene_thermal_2023, krishnamurthy_nondetection_2021} and TRAPPIST-1 e is likely to be an observing target in the future. Other rocky, presumably tidally locked planets whose rotation periods place them within the double jet regime in our parameter space include Teegarden's star b \citep{zechmeister_carmenes_2019} at 4.9 days and LP 890-9 c \citep{delrez_two_2022} at 8.5 days.

The single jet WOLF simulation presented in Section \ref{subsec:reg4} is likewise physically self-consistent, with the rotation period (~18 days), stellar spectrum, stellar constant, planet radius, and planet mass of Wolf 1061 c. This planet has not been modelled in GCM simulations before, but a frequently studied example of a planet usually simulated with a single superrotating jet is Proxima Centauri b \citep{boutle_exploring_2017, turbet_habitability_2016}, with a rotation period of 11.2 days. A transiting planet orbiting an M-class star with a similar stellar constant and rotation period as Wolf 1061 c is Kepler-1649 c \citep{vanderburg_habitable-zone_2020} at 19.5 days and a stellar constant of about 1.23 times that of Earth. A planet with a rotation period that places it in our single jet regime, but a stellar constant closer to that of TRAPPIST-1 e is GJ 1061 d \citep{anglada-escude_dynamically-packed_2013} at 13 days and a stellar constant about 0.69 times that of Earth.

As discussed in Section \ref{subsec:circreg}, however, models do not always predict the same circulation regime for a given rotation period. The ability to constrain a planet's circulation regime from observations would be a significant step forward in our understanding of exoplanet atmospheric dynamics. Our simulation results show distinct terminator asymmetry and limb silhouettes associated with each circulation regime. We propose that, for planets with optically thick haze, it may be possible to detect evidence of a circulation regime through separate atmospheric retrievals of the leading and trailing limb \citep{song_asymmetry_2021} or through retrieval of the planet's transmission string \citep{grant_transmission_2023}. We plan to test this approach in future work by producing simulated observations for our identified circulation regimes. However, simulating transmission spectra requires a larger atmospheric pressure range than included in our parameter space study, and the stratospheric circulation is also relevant to the haze distribution. It is possible that the low-haze troposphere of regime 2 (the transitional regime) is a result of particle trapping at the tropopause or in the stratosphere, which would in fact lead to a higher haze layer and poorer observational prospects. It may also be the case that the Rossby gyres extend into the stratosphere, again trapping haze particles at lower pressures. We will therefore extend the top of our model into the stratosphere for the purpose of simulating observations, as well as studying the potential stratospheric circulation.

\subsection{Limitations}\label{subsec:limitations}
Previous work reporting north-south asymmetry in GCM simulations of tidally locked planets has shown that the asymmetry may reverse periodically, i.e. the northern hemisphere may be warmer than the southern for some period of time, followed by a period in which the southern is warmer than the northern \citep{noda_circulation_2017}. As we use the average of one year for our analysis, our results do not reflect any such switching, and a long-term average (several times the period of the oscillation) could result in no differences between the northern and southern hemispheres. Since, however, transmission spectroscopy cannot distinguish latitudinal variations and any north-south asymmetry would therefore not affect observations, we chose a shorter averaging period to keep the simulation runtimes and output sizes at manageable levels. 

As stated in Section \ref{subsec:planets}, the model inputs for the WOLF simulations would likely lead to a runaway greenhouse regime in other, more sophisticated GCMs. \cite{haqq-misra_sparse_2022} found that ExoPlaSim underpredicts the stratospheric water vapour content compared to ExoCAM (their Table 3), an effect that is more pronounced in hotter simulations. \cite{paradise_exoplasim_2022} states as a possible explanation that ExoPlaSim treats water vapour as a trace gas, such that increases in water vapour pressure do not lead to increases in surface pressure and resulting effects on dynamics and climate. In addition, the absorptivity of water vapour in shortwave band 2 ( $\lambda > 0.75 \mu$m) relies on a parameterisation that is tuned to the solar spectrum and is likely less accurate for M-class stellar spectra. The WOLF simulations are therefore better seen as theoretical atmospheric states for warm temperate planets, rather than specific predictions for the stellar constant of Wolf 1061 c.

Our haze scheme is highly idealised and does not account for sinks (chemical sinks, wet deposition) aside from gravitational settling onto the surface, a particle size distribution (which may vary with time due to particle growth and loss), or radiative effects dependent on wavelength. The haze source is taken from 1-D modelling in \cite{arney_pale_2017} and prescribed within the model, not generated internally by self-consistent chemistry or haze microphysics. Accordingly, our results should be seen as first order estimates of the interaction between the atmospheric circulation and radiatively active haze particles. The strength of ExoPlaSim and our haze scheme is their computational efficiency and ability to map large parameter spaces and isolate relationships between a small number of factors. The free parameters in the haze scheme, including optical constants, particle size and density, and haze source strength and location, as well as the many model parameters relevant to the circulation and haze distribution (including stellar constant, stellar type, planet gravity, and others) leave a large parameter space still to be explored. We examine only one type of organic haze, which may vary in its optical properties from other laboratory haze analogues, as indeed real organic hazes may vary from planet to planet depending on factors such as trace gas abundances, stellar spectrum, and atmospheric temperature. Future work could, for example, study the effect of the organic haze data published in \cite{corrales_photochemical_2023} by way of comparison with our study.

We intend to follow up the present work with a smaller, more computationally intensive study of a select number of cases from our parameter space at a higher vertical and horizontal resolution and with a higher model top to investigate the transport of haze to and within the stratosphere and generate simulated transmission spectra. The source code of ExoPlaSim, including the haze scheme, is open source and freely available for further work by other researchers.

\section{Conclusion}\label{sec:conclusion}
We developed a haze module to simulate haze transport and radiative effects in a fast, open source intermediate complexity general circulation model, ExoPlaSim. We used the model to study the 3-D haze distribution for a tidally locked Earth-like (TRAPPIST-1 e-based) and super-Earth-like (Wolf 1061 c-based) planet with rotation periods ranging from 0.25 to 30 days in idealised simulations. The aim of our study was to identify characteristic haze distribution regimes which could potentially be distinguished by observations of the planetary limb. Our simulations fall into three primary circulation regimes, which have also been identified in previous literature: a banded regime (rapid rotators), a double jet regime (intermediate rotation periods), and a single jet regime (slow rotators). The banded regime tends to have haze particles well-mixed throughout the atmosphere, while in the other two regimes, haze collects either inside or around the Rossby gyres associated with the stationary Rossby waves typical of the atmospheric circulation of tidally locked planets. As the Rossby gyres are associated with large amounts of haze, their location--on the planetary limb or on the nightside--has a controlling effect on the haze differential optical depth at the terminator. Our simulations in the double jet and single jet regimes also show east-west terminator asymmetry, which could be observable. Overall, the simulations in our study with the lowest differential optical depths at the top of the atmosphere were the slowly rotating super-Earths, in which haze particles settled deeper into the atmosphere around the nightside Rossby gyres. Our chosen haze optical constants resulted in a haze layer which had little effect on the surface temperature of the planet or feedback on the general circulation. The organic hazes in our simulations therefore act largely as passive tracers; however, this finding cannot be generalised to other hazes or laboratory haze analogues, which must be examined individually.

Our model's computational efficiency allows us to explore a larger parameter space than is feasible with more complex GCMs. We intend to follow this work with a smaller set of higher resolution and higher model top simulations selected from the parameter space described in our results in Section \ref{sec:results}. This will allow us to explore the impact on the haze distribution of including the stratospheric circulation and to generate simulated transmission spectra and transmission strings. 

\section{Code availability} \label{sec:codeavail}
The release of ExoPlaSim including haze transport and radiative transfer is available at \url{https://doi.org/10.5281/zenodo.8410220}.
%% Also note that the akcnowlodgment environment does not support long amounts of text. If you have a lot of people and institutions to acknowledge, do not use this command. Instead, create a new \section{Acknowledgments}.
\section{Acknowledgments}
We acknowledge the funding and support provided by the Edinburgh Earth, Ecology, and Environmental Doctoral Training Partnership and the Natural Environment Research Council [grant number NE/S007407/1].
\vspace{5mm}


\software{We used MiePython \citep{prahl_miepython_2023} to calculate extinction and scattering efficiencies based on the haze complex refractive index.}

\appendix 
\section{} \label{Appendix}

\subsection{Gravitational settling scheme} \label{subsec:gsettle}

 ExoPlaSim includes a gridspace tracer transport module with a flux-form semi-Lagrangian transport scheme based on \cite{lin_multidimensional_1996} and \cite{lin_class_1994} that can be used to calculate the passive transport of gas-phase tracers. Vertical transport is calculated by a piecewise parabolic method as described in \cite{colella_piecewise_1984}. To represent solid-phase particle transport, we repurposed the advection fluxes in the zonal, meridional, and vertical directions from the existing flux-form scheme and added an additional vertical flux from gravitational settling based on \cite{steinrueck_3d_2021} and \cite{parmentier_3d_2013}.

According to this parameterisation, the haze mass mixing ratio is described by the equation:
\begin{equation}
    \frac{D\chi}{Dt} = -g \frac{\partial(\rho_g \chi V_s)}{\partial p} + P + L,
\end{equation}
where $\frac{D}{Dt}$ is the total derivative, $\chi$ is the mass mixing ratio (kg/kg), \emph{g} is the gravitational constant (m/s$^2$), $\rho_g$ is the surrounding gas density (kg/m$^3$), $V_s$ is the terminal velocity of the falling particles (m/s), \emph{p} is the pressure (Pa), \emph{P} is the production term (kg/kg), and \emph{L} is the loss term (kg/kg). The first term on the right-hand side represents gravitational settling of a solid-phase particle under gravity. The terminal velocity is given by:
\begin{equation}
    V_s = \frac{2 \beta a^2 g (\rho_p - \rho_g)}{2 \eta},
\end{equation}
where \emph{a} is the particle radius (m), \emph{g} is the gravitational constant (m/s$^2$), $\rho_p$ is the particle density (kg/m$^3$), $\rho_g$ is the surrounding gas density (kg/m$^3$), $\beta$ is the Cunningham correction factor (dimensionless), and $\eta$ is the viscosity (Pa $\cdot$ s). The Cunningham factor is a correction to the Stokes' drag force to account for the case in which the mean free path of the gas is similar to the particle size. Stokes' law assumes a no-slip condition, but this assumption is not valid for small particles, which experience the surrounding gas as a non-continuous medium. In this case, drag is generated by particles interacting with the surrounding individual gas molecules rather than by the viscosity of the gas. The Cunningham factor accounts for this effect and is defined by:
\begin{equation}
    \beta = 1 + Kn(1.256 + 0.4e^{-1.1/Kn}),
\end{equation}
where \emph{Kn} is the Knudsen number, the ratio of the mean free path $\lambda$ of the gas to the particle size \emph{a}. The mean free path is calculated as:
\begin{equation}
    \lambda = \frac{k_b T}{\sqrt{2} \pi d^2 P},
\end{equation}
where $k_B$ is the Boltzmann constant, \emph{T} is the air temperature (K), \emph{d} is the molecular diameter of the surrounding gas (m), and \emph{P} is the air pressure (Pa). The viscosity is calculated using a parameterisation described by \cite{rosner_transport_1986}:
\begin{equation}
    \eta = \frac{5}{16} \frac{\sqrt{\pi m k_b T}}{\pi d^2} \frac{(k_b T/ \epsilon)^{0.16}}{1.22},
\end{equation}
where \emph{m} is the molecular mass (kg) and $\epsilon$ is the depth of the Lennard-Jones potential well. We use the following values for N$_2$ gas: m = $4.652 \times 10^{-26}$ kg, d = $3.64 \times 10^{-10}$ m, and $\epsilon = 95.5 k_b$ T. The parameterisation represents particles with a single, uniform radius.
 
\subsection{Haze radiative transfer scheme} \label{subsec:aerorad}

 ExoPlaSim's shortwave radiation module is based on the work of \cite{lacis_parameterization_1974} for the clear sky fraction, with cloud parameterizations from \cite{stephens_parameterization_1984} and \cite{stephens_radiation_1978} for the cloud cover fraction. The model uses a two-stream method to represent multi-scattering by clouds in two shortwave bands and a diffusivity factor of 0.5 for diffuse light. Cloud is treated as scattering but not absorbing in shortwave band 1 (blue), and as both scattering and absorbing in shortwave band 2 (red). We added a haze radiative transfer scheme based on the cloud scheme, with adaptations to include empirically measured optical data for hazes rather than internally calculated values from parameterisations as for the clouds. Haze radiative effects are included only for the clear sky fraction; in cloudy areas, cloud scattering and/or absorption is assumed to dominate the radiative transfer.

\subsubsection{Cloud scheme} \label{subsubsec:cloudscheme}
The cloud treatment for shortwave band 1 is given in the Planet Simulator Reference Manual. We do not reproduce it here, as our additions to code are based only on the cloud scheme for shortwave band 2. The cloud treatment in shortwave band 2, representing both scattering and absorption, is as follows:
\begin{equation}
    Tr_{C2} = \frac{4u}{R},
\end{equation}
where \emph{Tr} is the transmissivity through a layer of cloud and,
\begin{equation}
    u^2 = \frac{(1 - \omega_0 + 2 \beta_2 \omega_0)}{(1-\omega_0)},
\end{equation}
where $\omega_0$ is the single scattering albedo (the ratio of the scattering efficiency to the total extinction efficiency and $\beta_2$ is the backscatter ratio (the backscattered fraction of monodirectional incident radiation at a given zenith angle, calculated from a parameterisation based on fields within the model) for shortwave band 2, and
\begin{equation}
    R = (u+1)^2 exp(\tau_{eff}) - (u-1)^2 exp(-\tau_{eff}),
\end{equation}
where $\tau_{eff}$ is the effective optical depth. The value of $\tau_{eff}$ is calculated using a parameterisation based on the cloud liquid water path, the single scattering albedo, the backscatter ratio, and the cosine of the solar zenith angle taken from \cite{stephens_radiation_1978}. The PlaSim Reference Manual provides more detailed information about this parameterisation.

In addition, the reflectivity from a layer of cloud is given by:
\begin{equation}
    Re_{C2} = \frac{[exp(\tau_{eff}) - exp(-\tau_{eff})](u^2 - 1)}{R},
\end{equation}
where \emph{u}, \emph{R}, and $\tau_{eff}$ are as given above. The transmissivity \emph{Tr} and the reflectivity \emph{Re} are unitless and are expressed as fractions with a value between 0 and 1. Detailed derivations of all the above equations are given in the PlaSim Reference Manual and in \cite{stephens_radiation_1978}.

\subsubsection{Haze scheme} \label{subsubsec:hazescheme}
The haze treatment in both shortwave bands 1 and 2 uses the same equations for \emph{Tr}, \emph{Re}, \emph{u} and \emph{R} as the cloud scheme above. The haze optical depth of a haze layer is given as:
\begin{equation}
    \tau_N = N Q_{ext} \pi r^2 \delta z,
\end{equation}
where \emph{N} is the number density in particles per $m^3$, $Q_{ext}$ is the extinction efficiency given in Table \ref{tab:mieqs}, \emph{r} is the particle radius in meters, and $\delta z$ is the layer thickness in meters.
The effective optical depth $\tau_{eff}$ is then given by:
\begin{equation}
    \tau_{eff} = \frac{[(1 - \omega_0)(1 - \omega_0 + 2 \beta \omega_0)]^{1/2} \tau_N}{\mu_0},
\end{equation}
in the formulation given by \cite{stephens_radiation_1978} for an absorbing and scattering medium, where $\omega_0$ is the single scattering albedo (the ratio of the scattering efficiency to the extinction efficiency), $\beta$ is the backscatter ratio (calculated for the haze as $dQ_{back}/d\Omega$ divided by $Q_{scat}$, the ratio of the backscattering efficiency per steradian to the scattering efficiency), and $\mu_0$ is the cosine of the solar zenith angle. Instead of being determined by parameterisations within the model, the values of $Q_{ext}$, $dQ_{back}/d\Omega$, and $Q_{scat}$ are read in from an external data file after being pre-calculated from the haze complex refractive index and are then used to calculate $\omega_0$ and $\beta$. Whereas the backscatter ratio in the cloud scheme is based on a parameterisation that takes into account varying zenith angle, in the absence of such a parameterisation we have kept the backscatter ratio identical for all zenith angles.

\subsection{Mie efficiency calculations}\label{subsec:miecalcs}

We use the Python package MiePython \citep{prahl_miepython_2023} to calculate the Mie extinction, scattering, and backscattering efficiencies. MiePython uses algorithms based on \cite{wiscombe_improved_1980} to calculate these efficiencies. The online documentation for the MiePython package contains detailed descriptions of the formulas used and examples of calculations, as well as validations against test cases reported in \cite{wiscombe_improved_1980} and \cite{bohren_absorption_1998}. We reproduce below the formulas used by the MiePython package to calculate the extinction, scattering, and backscattering efficiencies:

Extinction efficiency:
\begin{equation}
    Q_{ext} = \frac{2}{x^2} \Sigma_{n=1}^\infty (2n +1) (|a_n|^2 + |b_n|^2)
\end{equation}

Scattering efficiency:
\begin{equation}
    Q_{scat} = \frac{2}{x^2} \Sigma_{n=1}^\infty (2n +1) \cdot Re \{a_n + b_n\}
\end{equation}

where in both cases \emph{x} is the size parameter of the particle, $\frac{2 \pi a}{\lambda_{vac}}$, with \emph{a} the particle radius and $\lambda_{vac}$ the wavelength of incoming light in a vacuum, and 

\begin{equation}
    a_n = \frac{[D_n(mx)/m + n/x] \psi_n(x) - \psi_{n-1}(x)}{[D_n(mx)/m + n/x] \xi_n(x) - \xi_{n-1}(x)}
\label{eqn:an}
\end{equation}
\begin{equation}
    b_n = \frac{[m D_n(mx) + n/x] \psi_n(x) - \psi_{n-1}(x)}{[m D_n(mx) + n/x] \xi_n(x) - \xi_{n-1}(x)}
\label{eqn:bn}
\end{equation}

where \emph{x} is again the size parameter, \emph{m} is the complex refractive index of the particle, and $\psi$ and $\xi$ are equivalent to:

\begin{equation}
    \psi_n(x) = x j_n(x)
\end{equation}
\begin{equation}
    \xi_n(x) = x j_n(x) + i x y_n(x)
\end{equation}

Here $j_n(x)$ and $y_n(x)$ are spherical Bessel functions. The factor $D_n$ in Equations \ref{eqn:an} and \ref{eqn:bn} is the logarithmic derivative computed using the continued fraction method developed by \cite{lentz_continued_1990}. As the description of this method is lengthy and is fully documented in \cite{prahl_miepython_2023} and \cite{lentz_continued_1990}, we refer the reader to these publications for a full derivation.

Finally, we calculate the backscattering efficiency using MiePython as:

\begin{equation}
    Q_{back} = \frac{1}{x^2} | \Sigma_{n=1}^\infty (2n + 1) (-1)^2 (a_n - b_n) |^2
\end{equation}

where $a_n$ and $b_n$ are defined as in Equations \ref{eqn:an} and \ref{eqn:bn} above.

The calculation of the backscattering efficiency by MiePython is based on a definition of the backscattering cross-section used in optics, which differs by a factor of $4 \pi$ from the definition used in meteorology. We quote from the American Meteorological Society's Glossary of Meteorology:

`` Backscattering cross-section:

For plane-wave radiation incident on a scattering object or a scattering medium, the ratio of the intensity scattered in the direction toward the source to the incident irradiance.

So defined, the backscattering cross section has units of area per unit solid angle, for example, square meters per steradian.

In common usage, synonymous with radar cross section, although this can be confusing because the radar cross section is $4 \pi$ times the backscattering cross section as defined in 1) and has units of area, for example, square meters.''

(American Meteorological Society, 2023: Backscattering cross section,

https://glossary.ametsoc.org/wiki/Backscattering\_cross\_section)

By this definition, MiePython computes the radar cross-section, which can in some cases be larger than the scattering cross-section. However, the radiative transfer scheme described in Appendix Section \ref{subsec:aerorad} requires as input the backscatter ratio, the fraction of incident monodirectional light scattered back towards the source at a given zenith angle. This ratio must be less than or equal to one and the backscattering cross-section must be less than or equal to the total scattering cross-section. To arrive at the meteorological definition of the backscattering cross-section, therefore, the value computed by MiePython must be divided by a factor of $4 \pi$. As the backscattering efficiency differs from the backscattering cross-section only by a factor of the geometric cross-section (constant for a given particle radius, as in our simulations), we take the backscattering efficiency calculated by MiePython and divide it by a factor of $4 \pi$ to obtain the values of the backscattering efficiencies given in Table \ref{tab:mieqs}. We therefore express the value in Table \ref{tab:mieqs} as $dQ_{back}/d\Omega$, the backscattering efficiency per steradian.

The Mie efficiencies input into the ExoPlaSim model need not be calculated using MiePython, but can be obtained by other methods chosen by the user. However, care should be taken that the backscattering efficiency used is equivalent to the American Meteorological Society definition and that its value does not exceed that of the scattering efficiency.

%\bibliography{references}{}
\bibliographystyle{aasjournal}



\end{document}